%% file: Sub_chain_beambook_12012021.tex
\documentclass[conference]{IEEEtran}

\usepackage{cite}
\usepackage{amsmath,amssymb,amsfonts}
\usepackage{algorithmic}
\usepackage{graphicx}
\usepackage{textcomp}
\usepackage{xcolor}
\usepackage{hhline}
\input{input.tex}

\begin{document}
%

\title{Sub-Chain Beam for mmWave Devices: A Trade-off between Power Saving and Beam Correspondence}
\author{\IEEEauthorblockN{Jianhua~Mo\IEEEauthorrefmark{1}, Daehee~Park\IEEEauthorrefmark{2}, Boon~Loong~Ng\IEEEauthorrefmark{1}, Vutha~Va\IEEEauthorrefmark{1}, Anum~Ali\IEEEauthorrefmark{1}, Chonghwa~Seo\IEEEauthorrefmark{2}, and~Jianzhong~Charlie~Zhang\IEEEauthorrefmark{1}
	}
\IEEEauthorblockA{\IEEEauthorrefmark{1}Standards and Mobility Innovation Laboratory, Samsung Research America, Plano, TX 75023, USA
	\\Email: \{jianhua.m, b.ng, vutha.va, anum.ali, jianzhong.z\}@samsung.com}
\IEEEauthorblockA{\IEEEauthorrefmark{2}Samsung Electronics, Suwon, Korea
	\\Email: \{daehee7.park, chseo\}@samsung.com
}
}
\maketitle
\begin{abstract}
Beam correspondence, or downlink-uplink (DL-UL) beam reciprocity, refers to the assumption that the best beams in the DL are also the best beams in the UL. This is an important assumption that allows the existing beam management framework in 5G to rely heavily on DL beam sweeping and avoid UL beam sweeping: UL beams are inferred from the measurements of the DL reference signals. Beam correspondence holds when the radio configurations are symmetric in the DL and UL. However, as mmWave technology matures, the DL and the UL face different constraints often breaking the beam correspondence. For example, power constraints may require a UE to activate only a portion of its antenna array for UL transmission, while still activating the full array for DL reception. Meanwhile, if the UL beam with sub-array, named as sub-chain beam in this paper, has a similar radiation pattern as the DL beam, the beam correspondence can still hold. This paper proposes methods for sub-chain beam codebook design to achieve a trade-off between the power saving and beam correspondence.
\end{abstract}


\section{Introduction}

In the millimeter wave band, antenna arrays are usually adopted by UE to generate high-gain beams than the single antenna, and thus resulting in higher SNR and throughput. For example, \cite{Mo_Jianhua_Access19, Raghavan_TCOM19, AlAmmouri_Access19, Zhao_Kun_Access19} presented possible antenna array setups for mmWave 5G phones, where $2 \times 1$, $4 \times 1$ or $2 \times 2$ arrays are adopted.
One of the key challenges for 5G and beyond is the UE power consumption \cite{Heng_COMM21}. The battery life and temperature control issue aggravates in the mmWave bands compared to the sub-6 GHz band.
When the phone is heating up quickly, one straightforward solution is to fall back to the sub-6 GHz and turn off the mmWave array. 
The LTE fallback is not desired in general. First, the maximum data rate decreases from Gbps to a few hundred  Mbps or less. Second, the frequent turn off/on of the mmWave antenna module incurs additional latency, power consumption, and even service disruption.

Instead of falling back to LTE, an alternative solution is to reduce the number of activated antenna elements.
%
%
Such kind of beam which only activates a part of the array is called  `sub-chain beam' in this paper, since the antennas on the same array are connected to the same RF chain. Meanwhile, the beam which activates the whole array is called `full-chain beam'. Note that the deactivation approach has been used to create wide beams \cite{He_Tong_MobileNetwAppl15, Xiao_Zhenyu_TWC16} in hierarchical codebook design, but in this paper, it is utilized to save the power and prevent overheating, instead of broadening the beamwidth.



\begin{figure}[t]
	\centering
	\subfigure[Downlink: UE activates all antennas for reception]{
		\includegraphics[width= 0.9\linewidth]{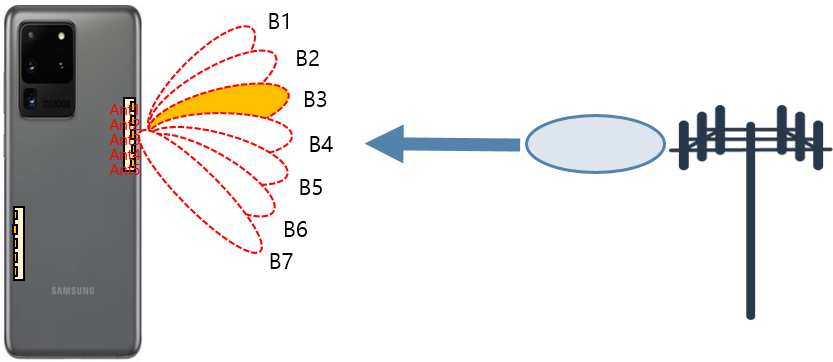}
		\label{fig:DL_5_chain}}
	\subfigure[Uplink: UE activates 3 antennas for transmission]{
		\includegraphics[width= 0.9\linewidth]{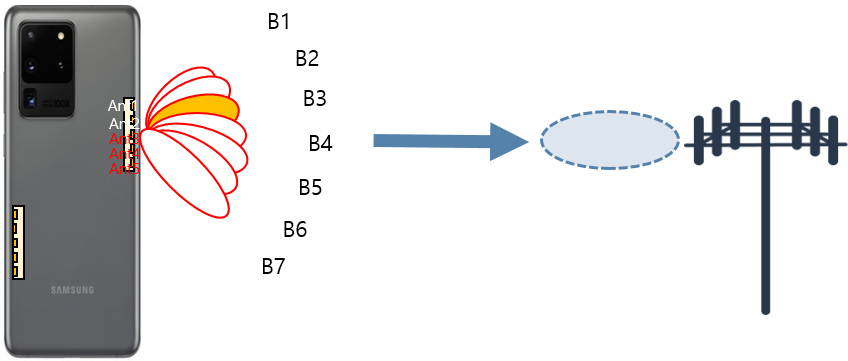}
		\label{fig:UL_3_chain}}
	\caption{UE receives with full-chain to maximize the beam gain and transmit with sub-chain to save the power and/or control the temperature. Dashed curves stand for Rx beams and solid curves stand for Tx beams.}
	\label{fig:DL_UL}
\end{figure}

An example of the sub-chain beam operation is shown in \figref{fig:DL_UL}, where UE activates only a portion of its antenna array for UL transmission, while still activating the full array for DL reception. This operation scheme is chosen since 1) the transmission consumes much more power than the reception, and 2) the downlink data rate requirement is usually higher than the uplink.

The 5G standard has defined the process of identifying and maintaining a suitable beam pair for the BS-UE link, which is known as beam management (BM) ~\cite{Eko_COMM18, Giordani2019Tutorial, Li_Access20, Ali_Access21}. 
The above operation scheme could destroy the DL-UL beam correspondence in 5G BM, which refers to the assumption that the best beams in the downlink direction are also the best beams in the uplink direction. 
DL-UL beam correspondence is an important design criterion, since an additional separate UL beam management procedure will be required if there is no DL-UL beam correspondence.


In this paper, we propose methods to design the sub-chain beam codebook to maximally maintain the beam correspondence. That means that if the full-chain Beam B3 in \figref{fig:DL_5_chain} is the best Rx beam in the downlink, the corresponding sub-chain Beam B3 in \figref{fig:UL_3_chain} should be the best Tx beam too. 

Another option of power saving is to scale down the transmission power level of all the antennas together. Although the total radiated power can be same between sub-chain and this option, the total power consumption is different. Activating a power amplifier (PA) requires a base power regardless of the power level. The option of reducing the power level still needs to turn on all the PAs, which means more base power consumption. In contrast, if we turn off some PAs (sub-chain beam), we can save not only the transmitted power but also the base power.

%
%

\textit{Notation:}  Bold uppercase letter $\mathbf{A}$ and bold lowercase letter $\mathbf{a}$ represents a matrix and a column vector, respectively. 
$\left(\cdot \right)^T, \left(\cdot\right)^*, \left(\cdot\right)^H$ denotes the transpose, conjugate and Hermitian of a vector or matrix, respectively. $\|\ba\|_0$ is the L0 norm of the vector $\ba$. 
$[\ba]_{m:n}$ stands for the a sub-vector of $\ba$ from the $m$-th entry to the $n$-th entry. 
$\mathbbm{1}\{\cdot\}$ is the indicator function.

\section{System Model}
\begin{table*}
	\small
	\centering
	\caption{An example of sub-chain beam codebook for a 1x5 single-polarization array with 2-bit phase shifters. 
	}\label{table:sub-chain beambook}
	\begin{tabular}{| c | c | c | c | c | c |}
		\hline
		\textbf{Beam index} & \textbf{Full-chain beam} 		& \textbf{4-Ant beam} 	& \textbf{3-Ant beam} 	& \textbf{2-Ant beam} 	& \textbf{1-Ant beam}\\ \hline
		$1$			& [1 1 1 1 1]$^T$				& [1  1 1  1 0]$^T$		& [0 0 1 1 1]$^T$			& [0 1 0 1 0]$^T$			& [0 0 0 0 1]$^T$ \\ \hline
		$2$			& [1 j -1 -j 1]$^T$				& [1 j 1 j 0]$^T$		& [j -1 -j  0 0]$^T$		& [1 j 0 0 0]$^T$			& [0 0 1 0 0]$^T$		\\ \hline
		$\cdots$	& $\cdots$ 						&$\cdots$				& $\cdots$					&$\cdots$					& $\cdots$	\\ \hline
		$K$			& [1 -1 1 -1 1]$^T$				& [1 j j 0 -1]$^T$		& [0 -1 -1  1 0]$^T$		& [0 -1 1 0 0]$^T$			& [0 1 0 0 0]$^T$\\ \hline
	\end{tabular}
\end{table*}

In this paper, we consider a UE equipped with 2 arrays on the left and right edge, respectively. In each array, there are $L$ dual-polarization patch antennas. 
The $2L$ antenna elements are denoted as 1V, 2V, $\cdots$, LV, 1H, 2H, $\cdots$, LH. The beamforming vector is thus defined as
\begin{align}
	\bw = \left[w_{1h},  w_{2h}, \cdots, w_{Lh}, w_{1v}, w_{2v}, \cdots, w_{Lv} \right]^T,
\end{align}
where the magnitude of the beamforming weights, i.e., $|w_{\ell}|$ ($1\leq \ell \leq 2L$), is either 0 or 1. In other words, the antenna is either on or off, without the capability of fine magnitude tuning. The phase of $\bw$ is also restricted to a few discrete values, since the finite-resolution phase shifters are used in practice. If $b$-bit phase shifters are adopted in the implementation, the constraint on a nonzero beamforming weight is $w_{\ell}^{2^b}=1$.

The radiation pattern of a practical mmWave antenna, combined with the impact of terminal housing, is highly irregular \cite{Hong_Wonbin_COMM14, Zhao_Kun_Access19, Zhang_Jin_TAP20}. We thus adopt the data-driven method for codebook design \cite{Mo_Jianhua_Access19}. The E-field response data of each antenna element is obtained from the simulation or measurement. Note that the terminal housing effect are incorporated in the E-field data. The beam pattern is calculated as,
\begin{align}
	B(\theta,\phi) = \bw^H \bM(\theta, \phi) \bw,
\end{align}
where $\bM(\theta, \phi)$ consists of the E-field response of each antenna in the direction $(\theta, \phi)$, and is usually a rank-2 matrix \cite{Mo_Jianhua_Access19}.

For the sub-chain beam, there could be additional design requirements based on the hardware implementation. In this paper, we consider a constraint that the number of activated antennas in the two polarizations are same, i.e., $\left\| \left[\bw\right]_{1:L} \right\|_0 = \left\| \left[\bw\right]_{L+1:2L} \right\|_0$, but the indices of the activated H and V-polarization antenna can be different. Note that it is straightforward to extend our design method to meet other constraints. For example, the indices of the activated antennas in the two polarizations should be same which means that $|w_{ih}| = |w_{iv}|$ for $1\leq i\leq L$.

An example sub-chain beam codebook is provided in Table \ref{table:sub-chain beambook} where $K$ is the codebook size. For simplicity, we only show single-array single polarization case in this table. When the UE switches between the full-chain operation to a sub-chain operation, or between two distinct sub-chain operations, it does not need to perform a new round of beam sweeping to identify the best beam if the beam correspondence has been maintained. Instead, it adopts the corresponding beam in the same row directly.

\section{Sub-chain beam codebook design}
There could be different metrics and methods to design the sub-chain beam codebooks per different requirements and UE operation procedures. In this paper, we consider three different metrics, which are called,
\begin{enumerate}
	\item `similarity score (Sim)', which puts great emphasize on the beam mapping accuracy when activating/deactivating antennas (i.e., each row of Table \ref{table:sub-chain beambook}); 
	\item `spherical coverage (SC)', which makes much account of the performance of fixed-antenna beam codebook (i.e., each column of Table \ref{table:sub-chain beambook});
	\item `beam correspondence spherical coverage (BC-SC)', which is a trade-off between the first two metrics.
\end{enumerate}

We denote the method maximizing these metrics as `Sim-Max', `SC-Max' and `BC-SC-Max', respectively. 

\subsection{Similarity Score Maximization (Sim-Max)} 
In the first method, the sub-chain beams are designed to resemble the full-chain beams. In other words, the radiation pattern of each sub-chain beam is designed to be similar to the corresponding full-chain beam (one-to-one mapping). There could be different measures of similarity of two beam patterns. In one approach, assuming that there is a set of $N_p$ uniformly distributed sampling points on the unit-sphere (e.g., Fibonacci grid \cite{Fibonacci_Swinbank06}),  the similarity score is defined as,
\begin{align}
	s_{ij} = & \frac{1}{\sum_{n=1}^{N_p} G_i^2 \left( \theta_n, \phi_n \right)} \sum_{n=1}^{N_p} G_i \left( \theta_n, \phi_n \right) B_j\left( \theta_n, \phi_n \right),
\end{align}
where $G_i \left(\theta, \phi \right)$ is the $i$-th full-chain beam pattern, and $B_j \left(\theta, \phi \right)$  is the sub-chain beam pattern. The term $\sum_{n=1}^{N_p} G_i^2 \left( \theta_n, \phi_n \right)$ is to normalize the score, such that the score between a beam and itself is one. 
The candidate sub-chain beam with the largest similarity score is chosen. The candidate sub-chain beam could be from an available pool of beams.
Alternatively, the sub-chain can be obtained by solving the following optimization problem,
\begin{align}
	\textbf{P1:} \quad \max_{\bw} \quad & \bw^H \left(\sum_{n=1}^{N_p} G_i \left( \theta_n, \phi_n \right) \bM \left( \theta_n, \phi_n \right) \right) \bw \\
	s.t., \quad  & w_{\ell} = 0 \text{ or } w_\ell^{2^b}=1, \quad \forall \ell, \label{P1_PS} \\
	& \|[\bw]_{1:L} \|_0 = L_A, \label{P1_h}\\
	& \|[\bw]_{L+1:2L} \|_0 = L_A, \label{P1_v}
\end{align}
where $L_A$ ($1\leq L_A < L$) is the number of activated antennas per polarization.

The discrete phase constraint \eqref{P1_PS} and $\text{L0}$ norm constraints \eqref{P1_h} \eqref{P1_v} in P1 are both non-convex, making it difficult to solve P1. Since the array size $L$  is usually small for the UE, we can exhaustively try all the possible activation of the antennas, and solve ${L \choose L_A}^2$ problems separately. For a given activation without $\text{L0}$ norm constraints, P1 can be solved efficiently by an iterative algorithm which optimizes the phase cyclically \cite[Algorithm 3]{Mo_Jianhua_Access19}. Although the iterative algorithm only provides a local optimum, we can run it multiple times with different initial beams, and select the best local optimum as the final solution. Since there are still $2^{2bL_A}$ possible beams given an activation, the iterative algorithm has much lower complexity than the exhaustive search when $b$ and $L_A$ is not too small (e.g., $b=5$ and $L_A=4$).

\subsection{Spherical Coverage Maximization (SC-Max)} In this method, the sub-chain codebook is designed to maximize the spherical coverage. There is no consideration on the one-to-one mapping between the sub-chain and full-chain beams, which implies that a fresh beam sweeping would be needed to determine the best beam if the UE switches from full-chain to sub-chain codebook (or switches between two sub-chain codebooks). This design could be adopted if the UE chooses to operate with same number of antennas for Tx and Rx, and thus DL-UL beam correspondence is maintained as in the conventional full-chain Tx-Rx case.

The codebook is designed to maximize the average beam  gain over the whole sphere. The optimization problem is as follows,
\begin{align}
	\textbf{P2:} \max_{\{\bw_k, 1\leq k \leq K\}} \quad & \frac{1}{N_p} \sum_{n=1}^{N_p} \left( \max_{k} \quad \bw_k^H  \bM \left( \theta_n, \phi_n \right)  \bw_k \right) \\
	s.t., \quad  & \left[\bw_k \right]_{\ell} = 0 \text{ or } \left[\bw_k \right]_{\ell}^{2^b}=1, \quad \forall \ell, \forall k \\
	& \left\| \left[\bw_k \right]_{1:L} \right\|_0 = L_A, \quad \forall k\\
	& \left\| \left[ \bw_k \right]_{L+1:2L} \right\|_0 = L_A, \quad \forall k.
\end{align}
The K-Means algorithm \cite{Mo_Jianhua_Access19} is adopted to solve P2. It iterates between two steps: i) assign each direction to the beam providing the largest gain, ii) optimize the beams to serve the assigned directions. In the beam optimization step, we solve ${L \choose L_A}^2 K$ problems by exhausting all the possible antenna activation.

\subsection{Beam Correspondence Spherical Coverage Maximization (BC-SC-Max)} In the third method, the sub-chain beams are designed to maximize the radiation pattern over the full-chain beam’s coverage region, which is a sub-region of the whole unit-sphere. A fresh beam sweeping is not necessary in this option since the sub-chain beam is designed to cover an angular region similar to the full-chain beam. Therefore, for a given channel, the sub-chain beam is very likely to be the best one if the corresponding full-chain beam is the best. 
The design procedure is as follows.
\begin{enumerate}
	\item Partition the unit-sphere into the coverage regions of the full-chain beams. If there are $K$ beams, we have $\mathcal{D}_1 \cup \mathcal{D}_2 \cup \cdots \mathcal{D}_k = \left\{(\theta, \phi)|0^\circ \leq \theta \leq 180^\circ, 0^\circ \leq \phi < 360^\circ  \right\}$ and $\mathcal{D}_i \cap \mathcal{D}_j = \emptyset, i\neq j$ where $\mathcal{D}_k$ is the disjoint angular region covered by the full-chain beam $\tilde{\bw}_k$,
	\begin{align}
			\mathcal{D}_k = \left\{(\theta, \phi) \bigg| k=\argmax_{1\leq i\leq K} \tilde{\bw}_i^H \bM(\theta, \phi) \tilde{\bw}_i \right\}.
	\end{align}
	\item For each angular region covered by the full-chain beam, design the best sub-chain beam by solving the following problem,
	\begin{align}
		\textbf{P3:} \quad \max_{\bw} \quad & \bw^H \left( \sum_{(\theta_n, \phi_n)\in \mathcal{D}_k}  \bM (\theta_n, \phi_n) \right) \bw \\
		s. t. & \quad \eqref{P1_PS}, \eqref{P1_h}, \eqref{P1_v}.
	\end{align}
\end{enumerate}

The optimization problem $\text{P3}$ is similar to $\text{P1}$ and thus the same iterative algorithm is adopted with two minor modifications. First, the summation in $\text{P1}$ is over the whole unit-sphere, but $\text{P3}$ is only over an angular region. Second, the summation in $\text{P1}$ is weighted by the full-chain beam pattern while there is no weights in P3. Roughly speaking, when generating sub-chain beam resembling the full-chain beam, SC-Max method examines the similarity over the whole sphere, while BC-SC-Max method only considers the main lobe region.   

\section{Simulation Results}

\begin{figure*}[th]
	\centering
	\subfigure[5-Ant]{
		\includegraphics[width=0.18 \linewidth]{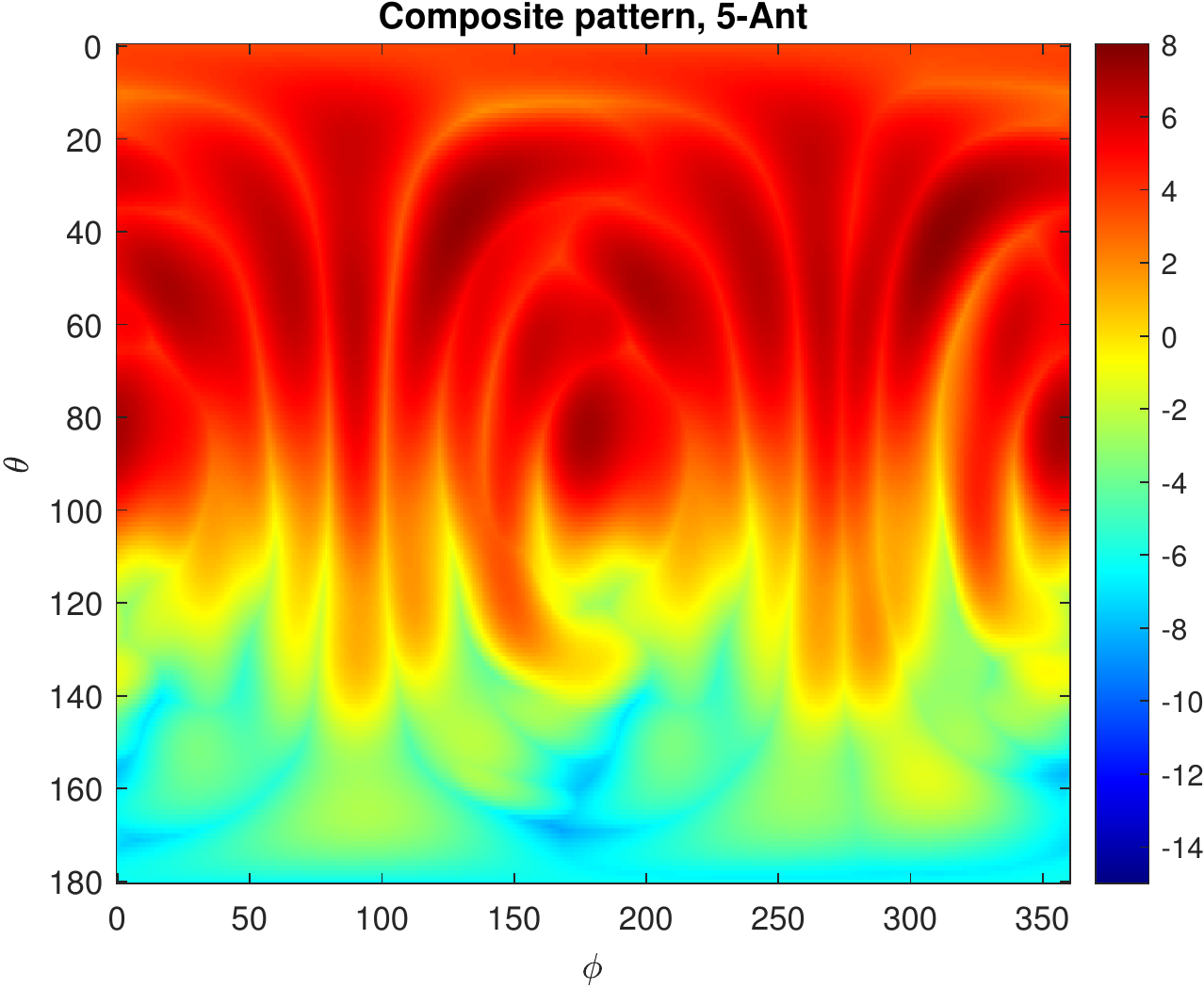}
		\label{fig:5_Ant_Pattern}}
	\subfigure[4-Ant]{
		\includegraphics[width=0.18 \linewidth]{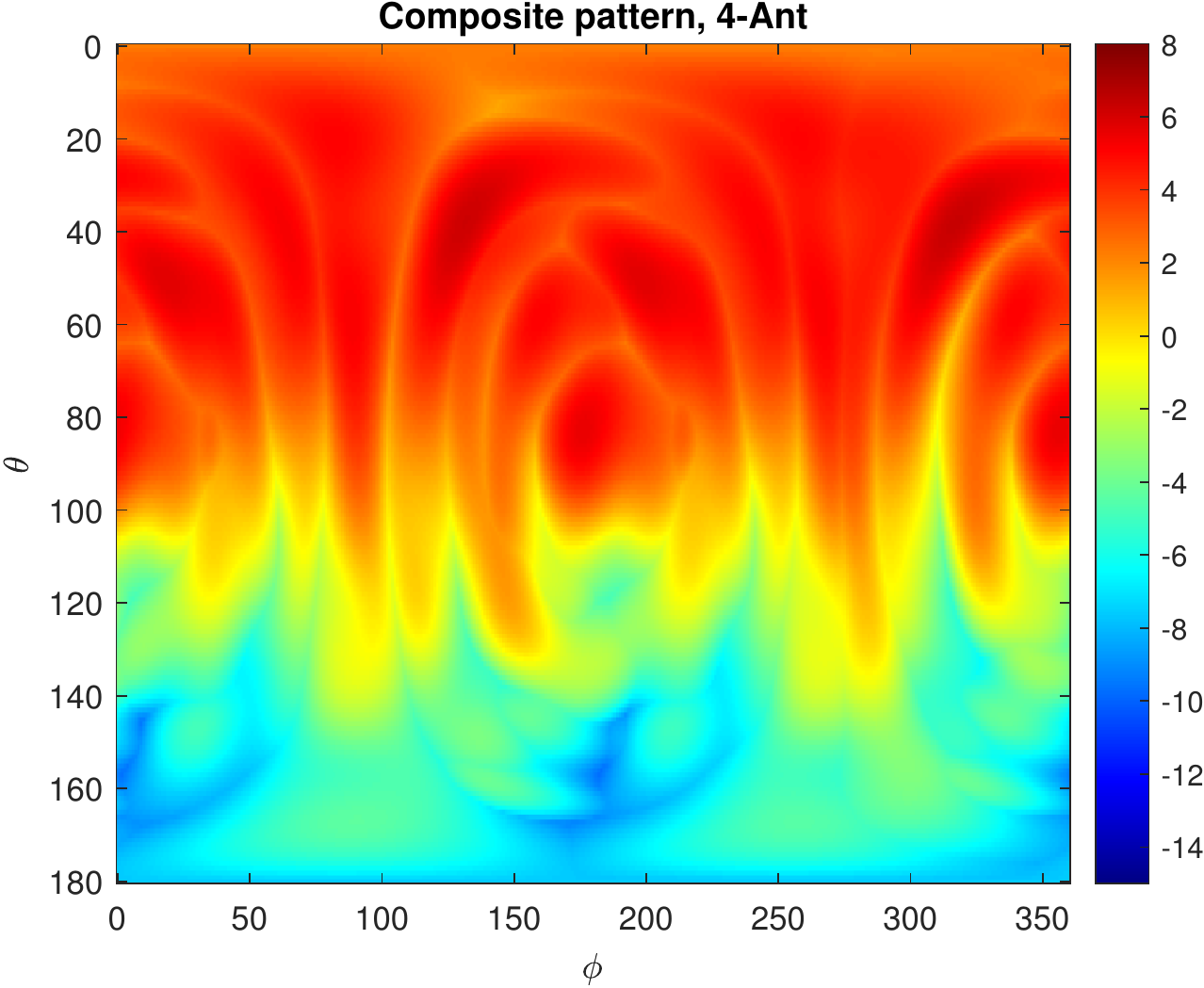}
		\label{fig:4_Ant_Pattern}}
	\subfigure[3-Ant]{
		\includegraphics[width=0.18 \linewidth]{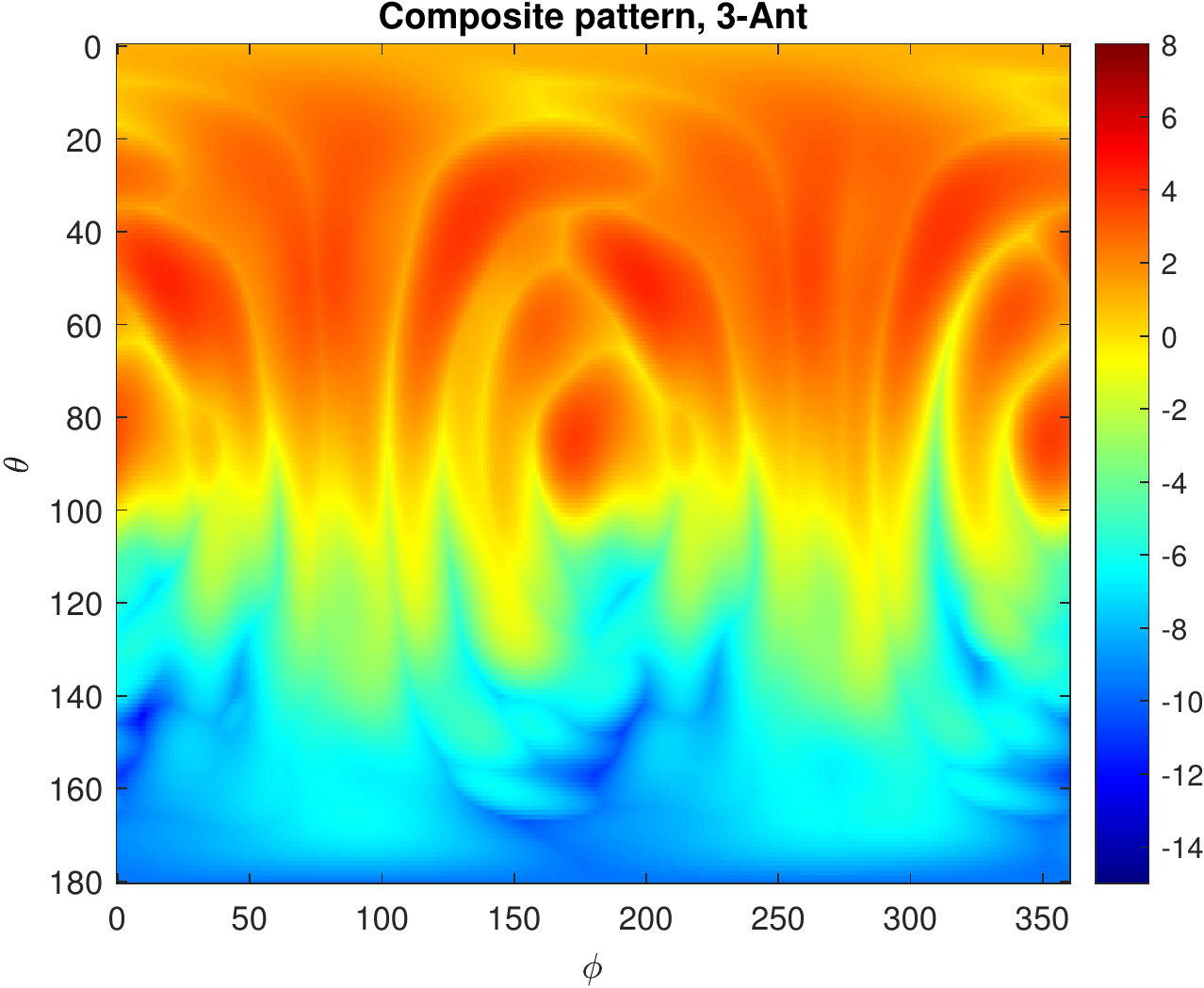}
		\label{fig:3_Ant_Pattern}}
	\subfigure[2-Ant]{
		\includegraphics[width=0.18 \linewidth]{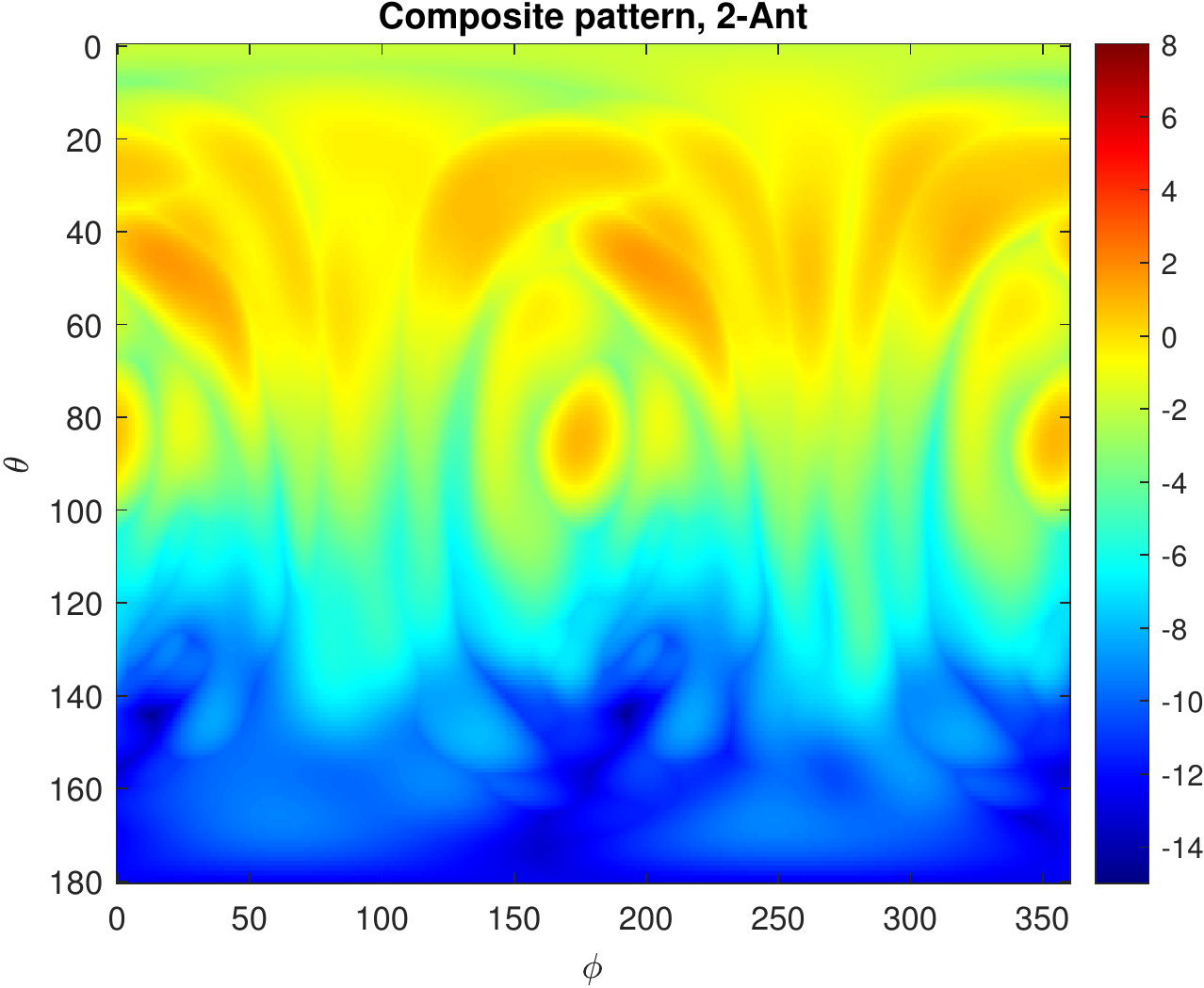}
		\label{fig:2_Ant_Pattern}}
	\subfigure[1-Ant]{
		\includegraphics[width=0.18 \linewidth]{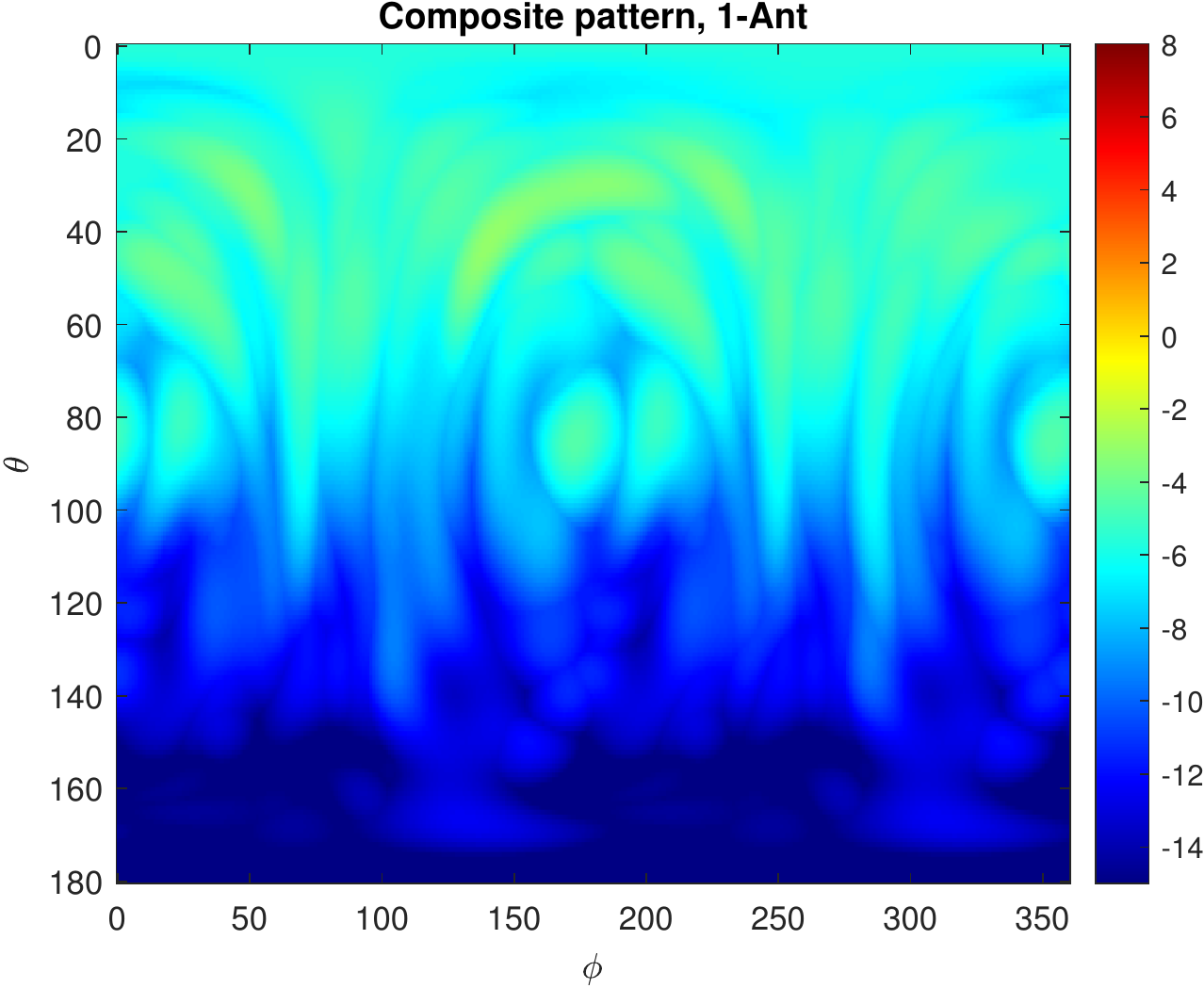}
		\label{fig:1_Ant_Pattern}}
	\subfigure[5-Ant]{
		\includegraphics[width=0.18 \linewidth]{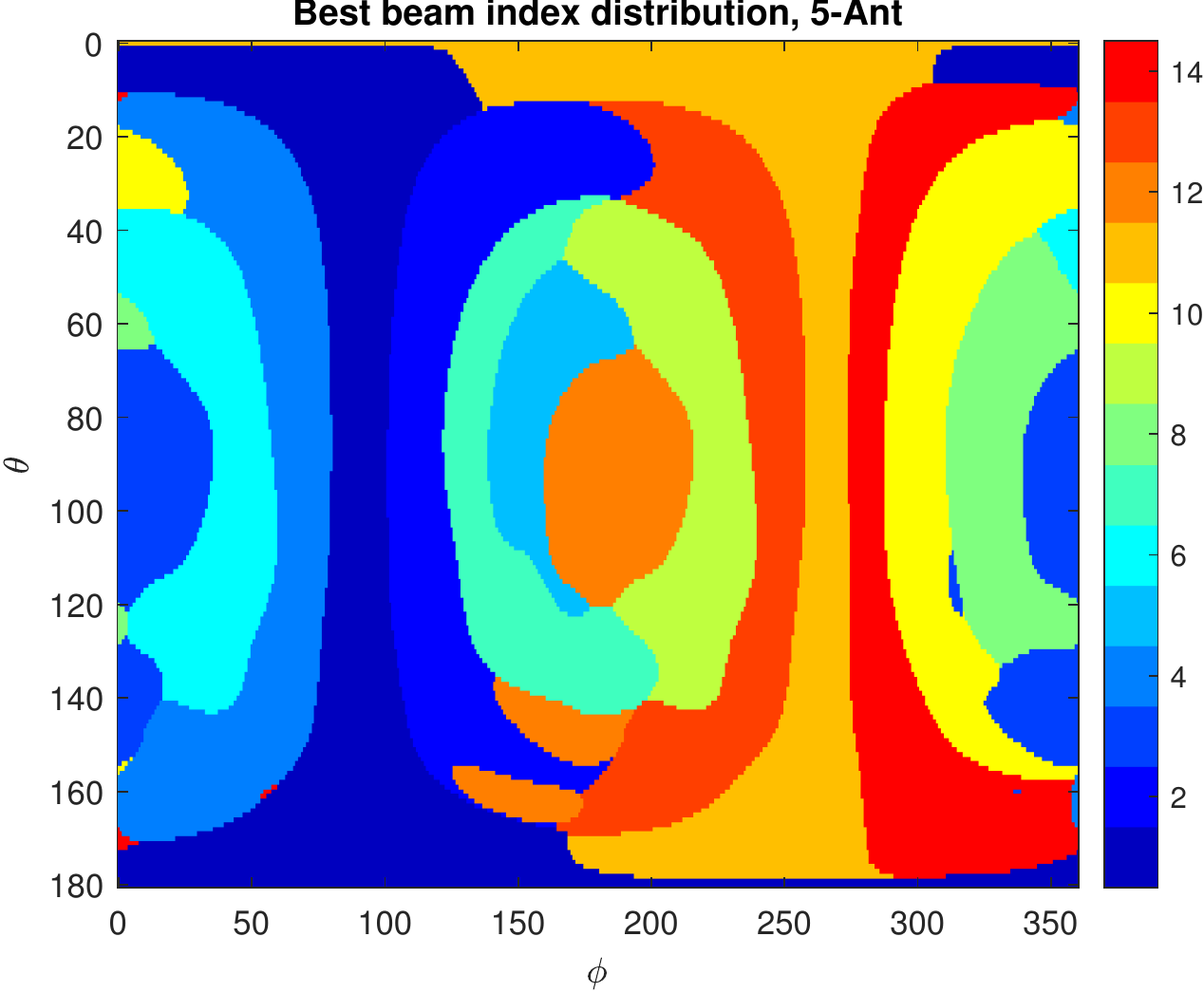}
		\label{fig:5_Ant_BestBeamIdx}}
	\subfigure[4-Ant]{
		\includegraphics[width=0.18 \linewidth]{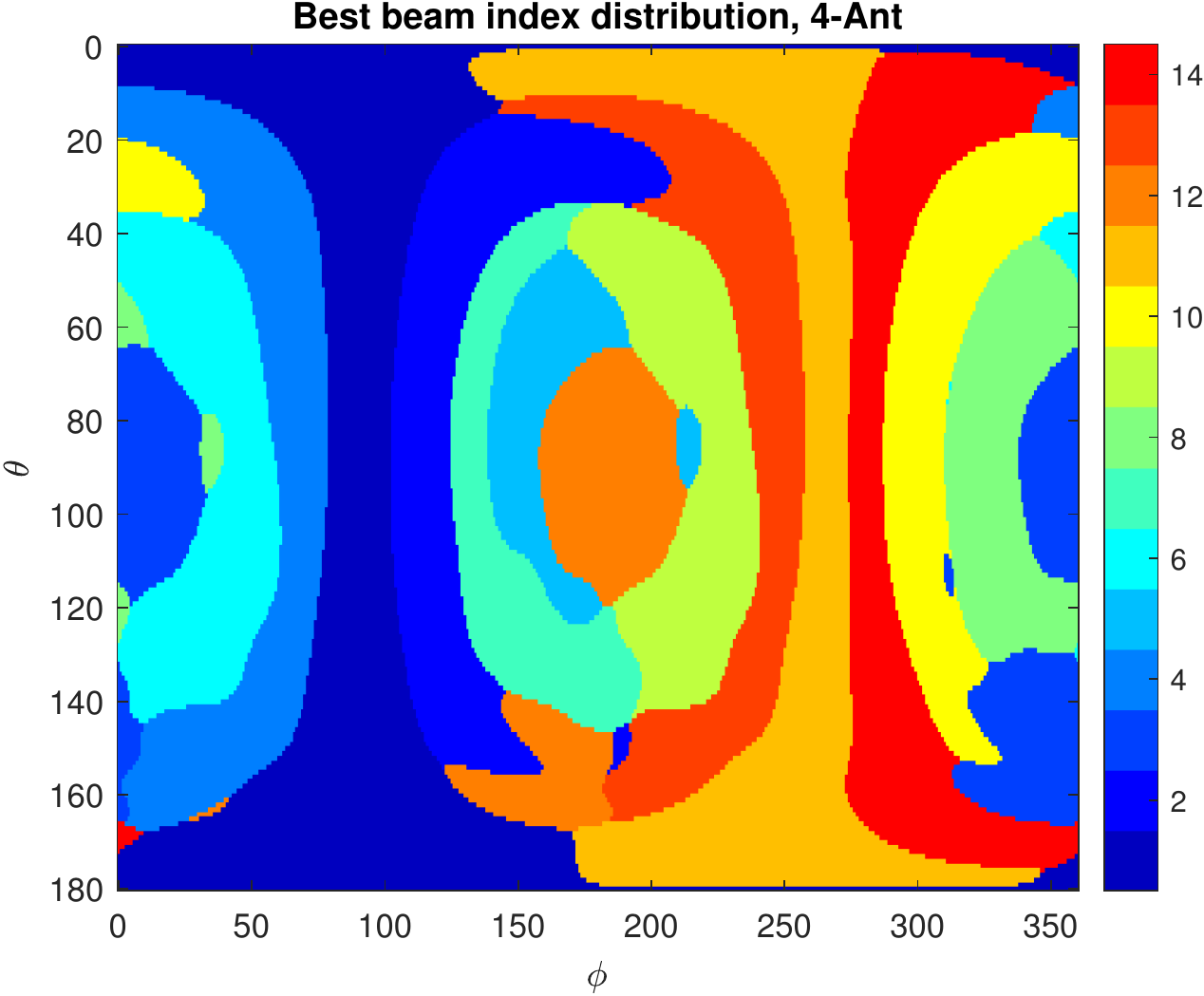}
		\label{fig:4_Ant_BestBeamIdx}}
	\subfigure[3-Ant]{
		\includegraphics[width=0.18 \linewidth]{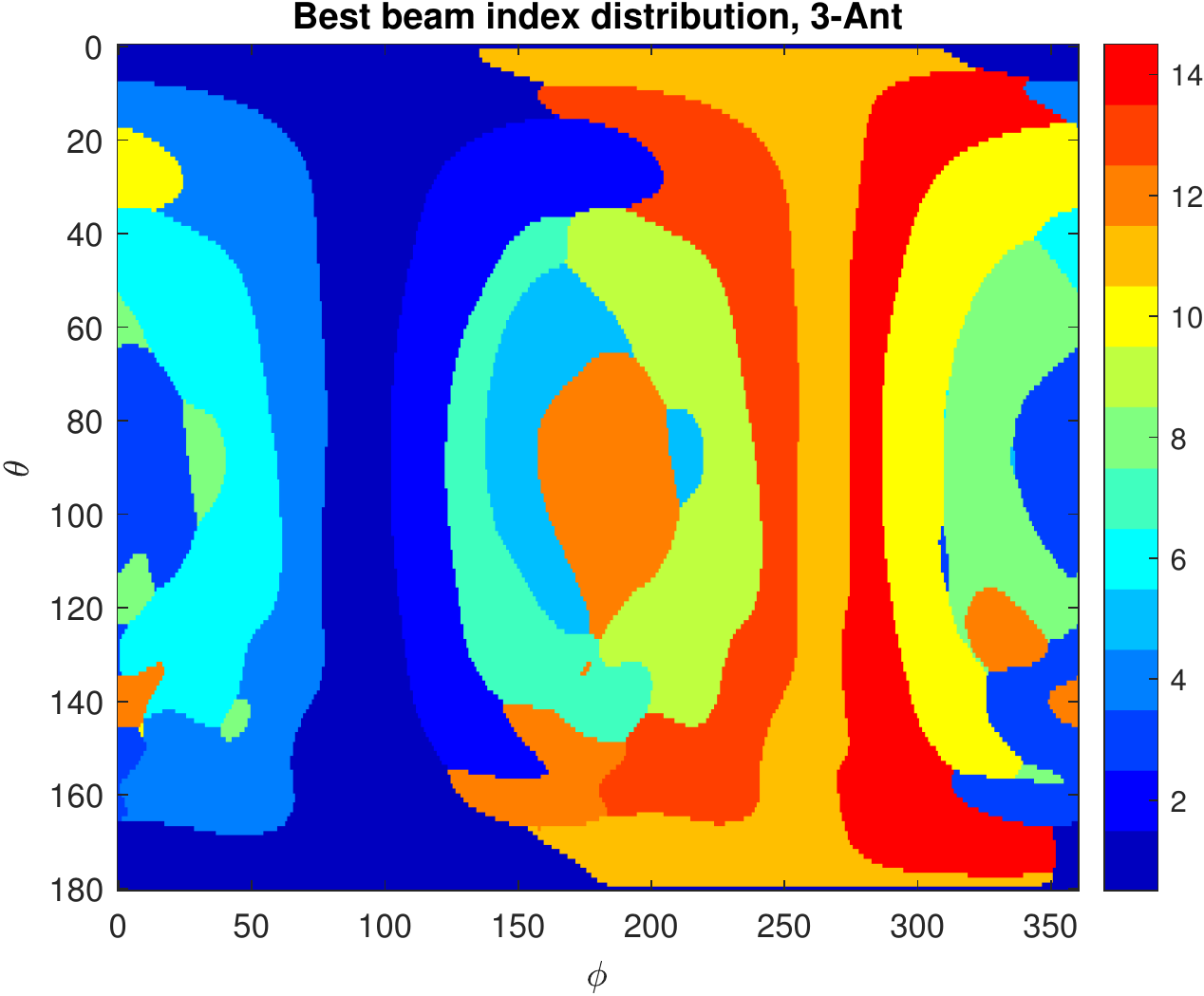}
		\label{fig:3_Ant_BestBeamIdx}}
	\subfigure[2-Ant]{
		\includegraphics[width=0.18 \linewidth]{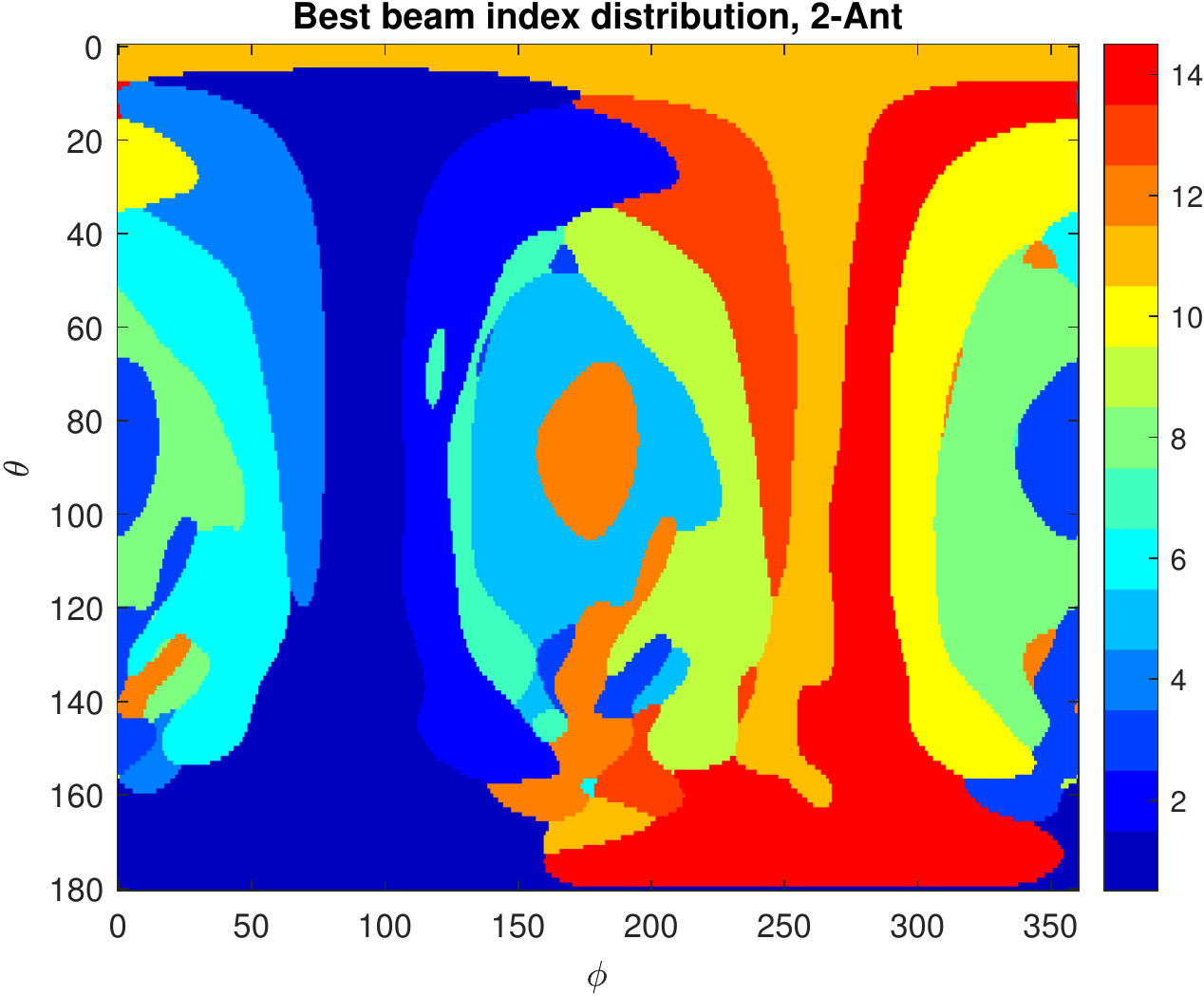}
		\label{fig:2_Ant_BestBeamIdx}}
	\subfigure[1-Ant]{
		\includegraphics[width=0.18 \linewidth]{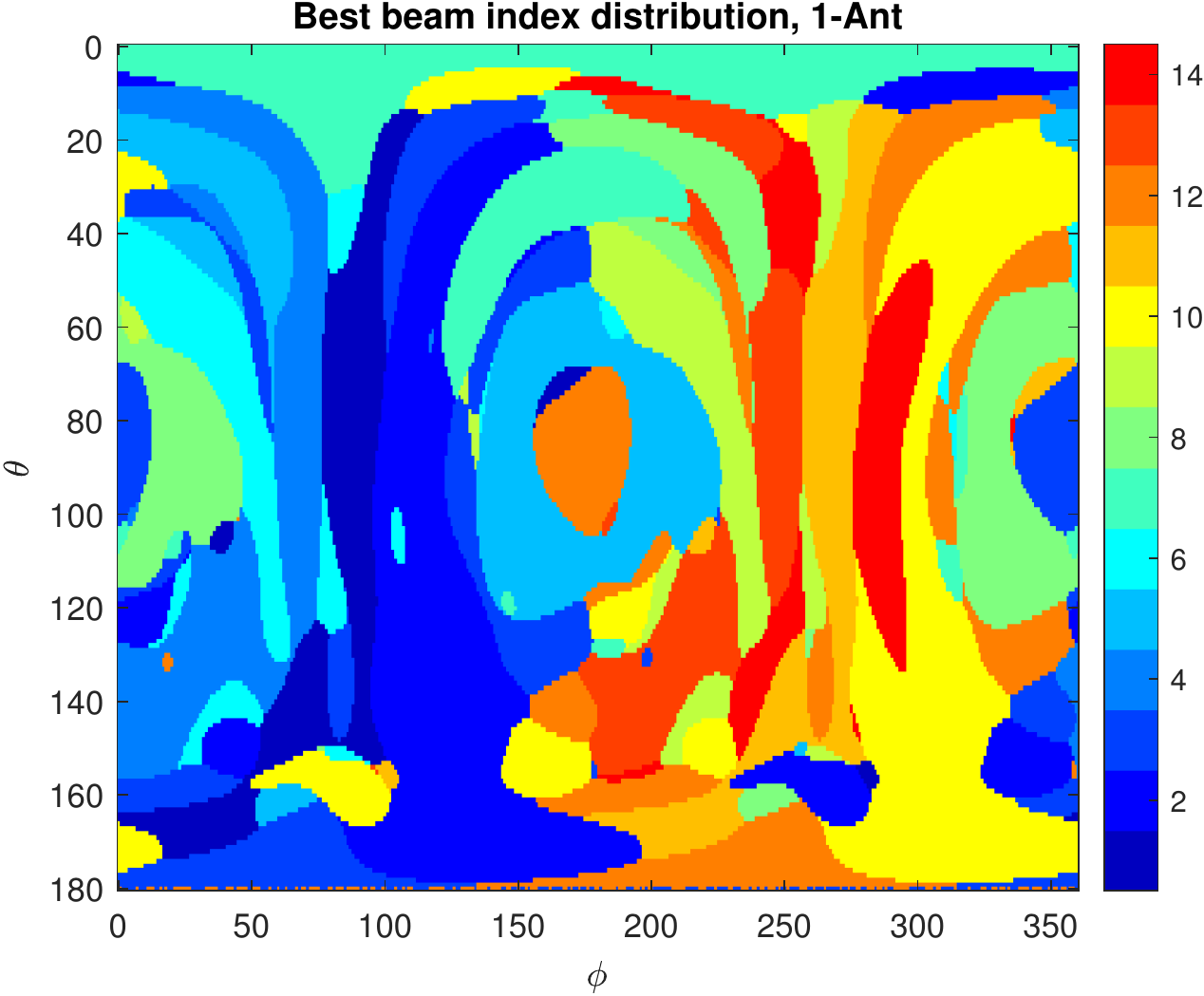}
		\label{fig:1_Ant_BestBeamIdx}}
	\caption{The pattern and the best beam index of the full-chain, and sub-chain codebooks generated by the Sim-Max method.}
	\label{fig:Sim_Max}
\end{figure*}

\begin{figure*}[th]
	\centering
	\subfigure[5-Ant]{
		\includegraphics[width=0.18 \linewidth]{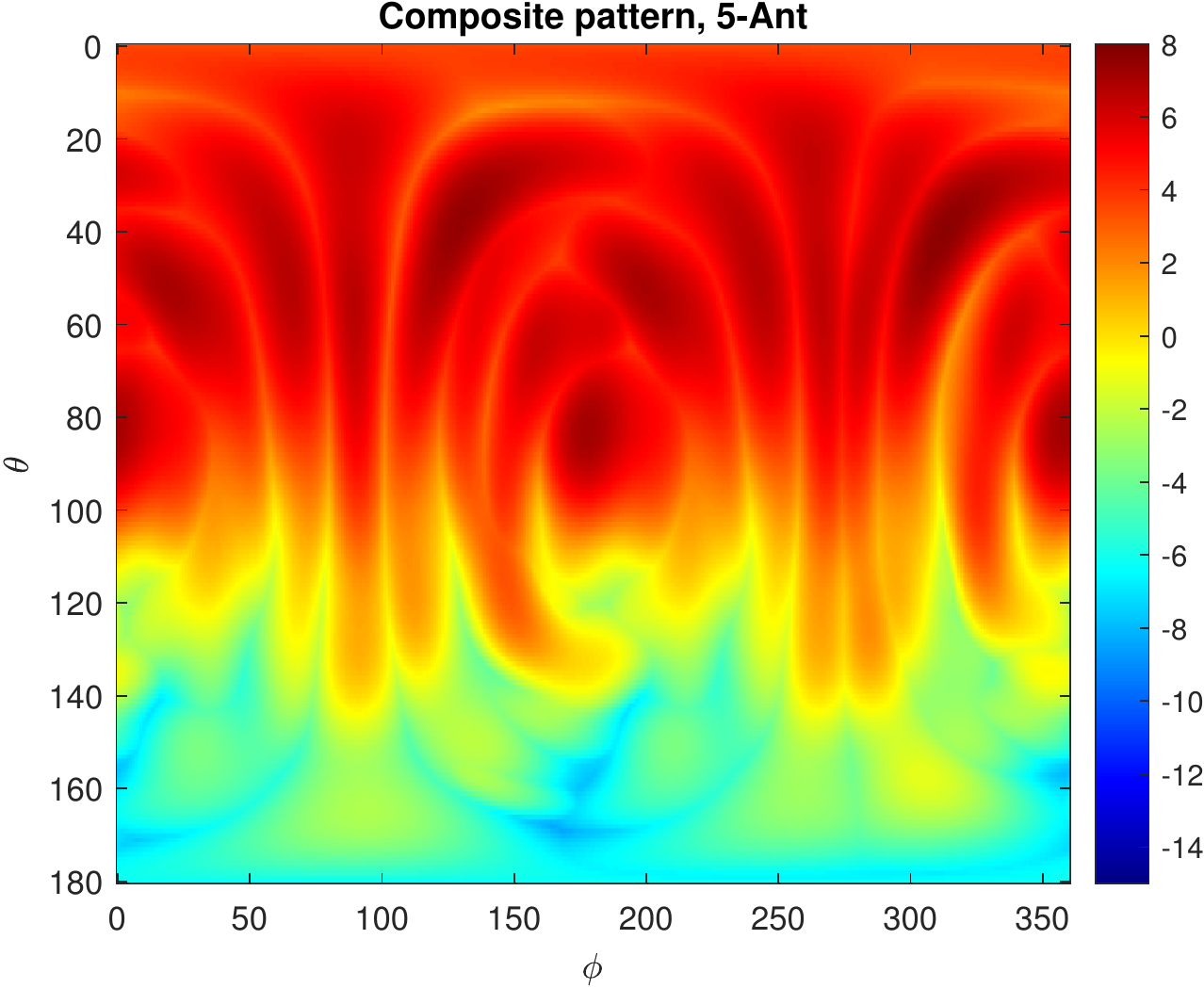}
	}
	\subfigure[4-Ant]{
		\includegraphics[width=0.18 \linewidth]{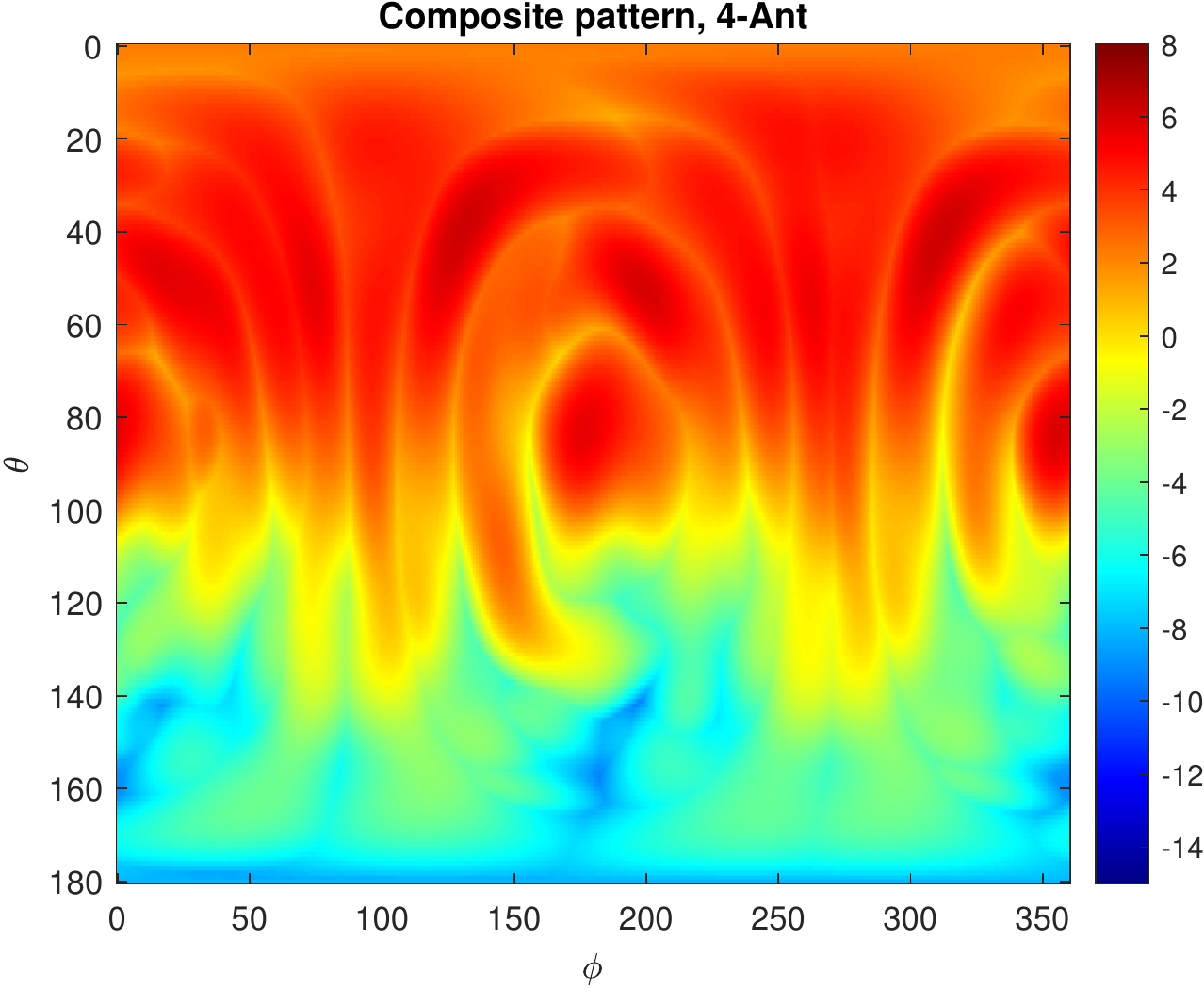}
	}
	\subfigure[3-Ant]{
		\includegraphics[width=0.18 \linewidth]{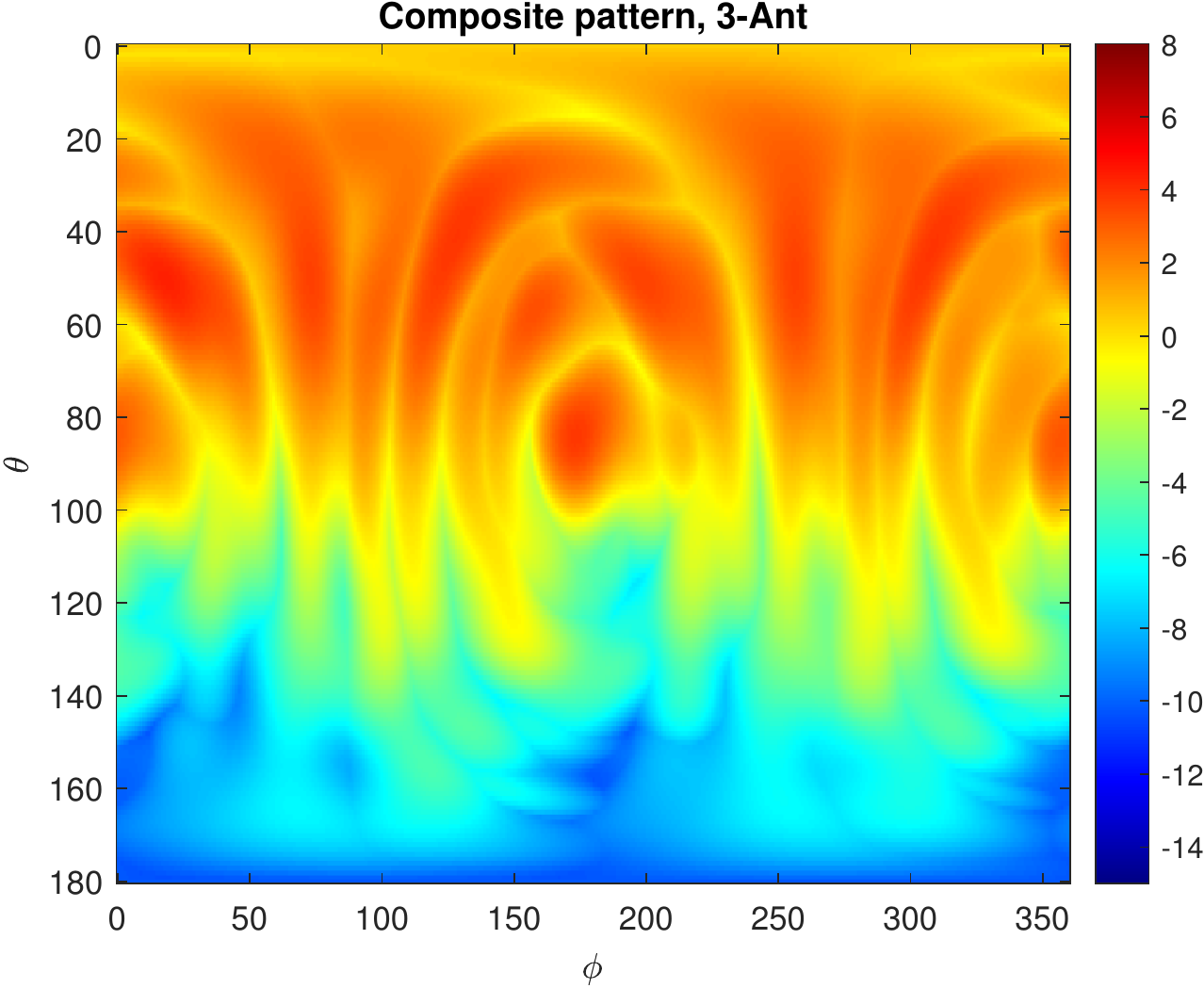}
	}
	\subfigure[2-Ant]{
		\includegraphics[width=0.18 \linewidth]{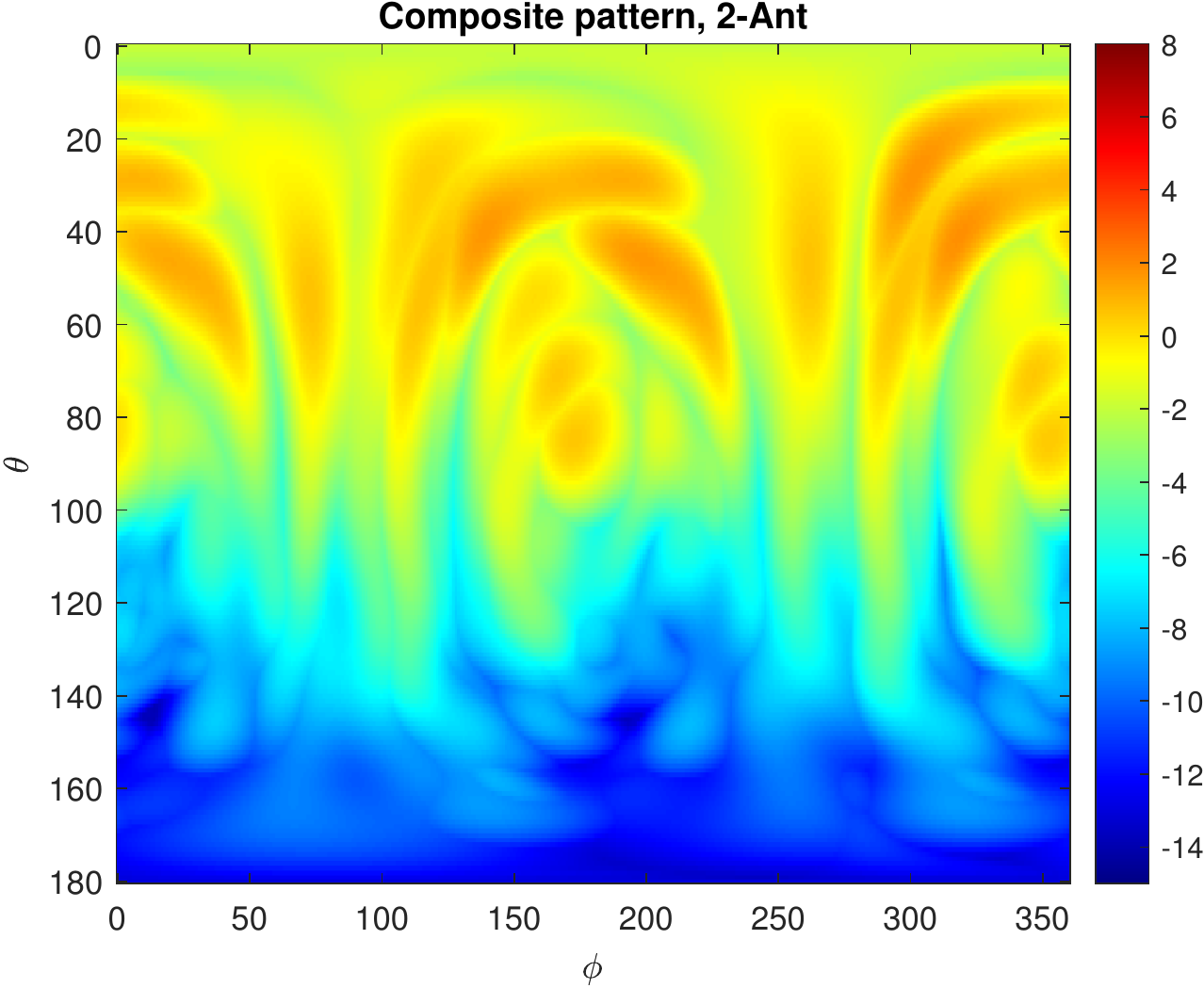}
	}
	\subfigure[1-Ant]{
		\includegraphics[width=0.18 \linewidth]{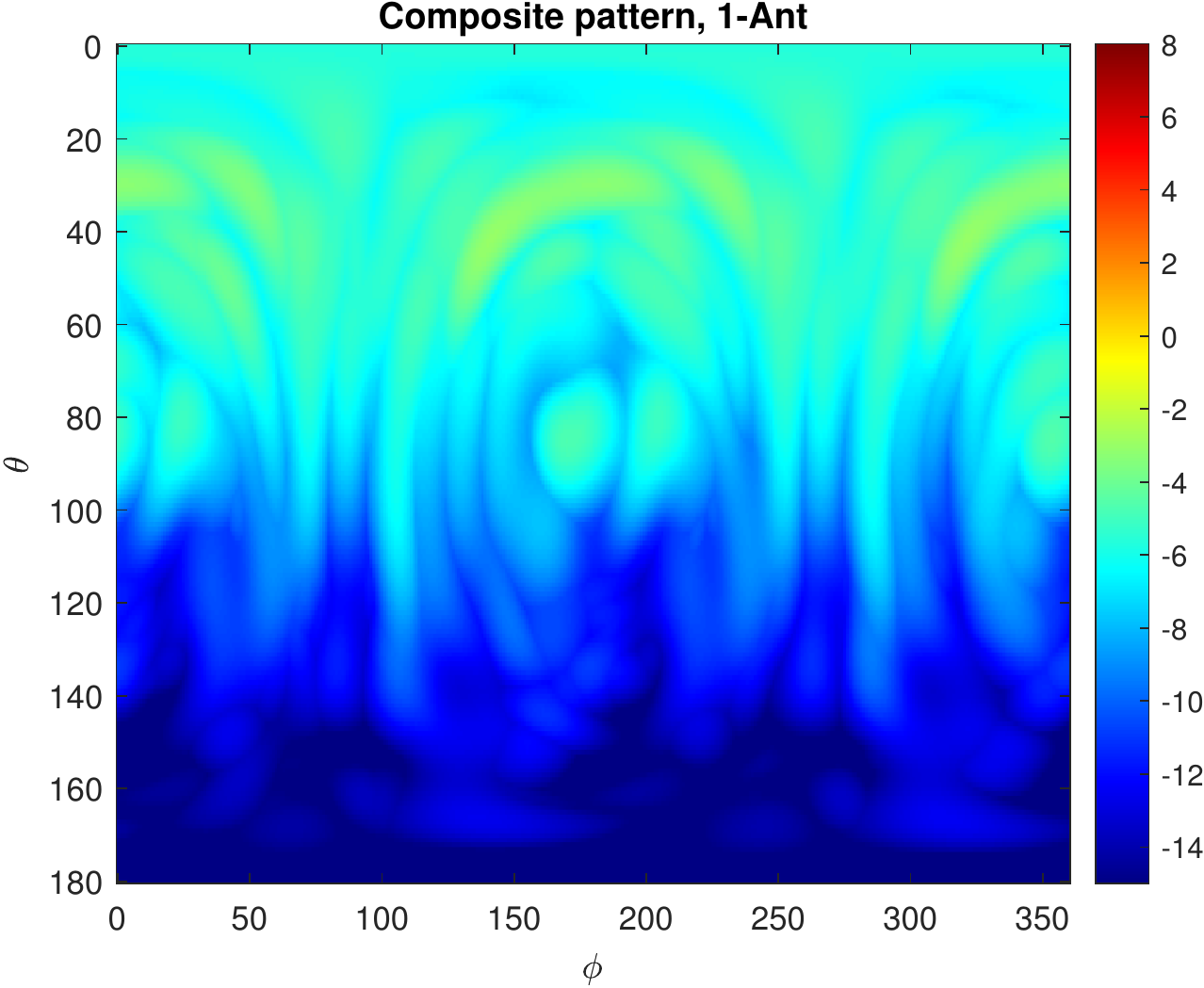}
	}
	\subfigure[5-Ant]{
		\includegraphics[width=0.18 \linewidth]{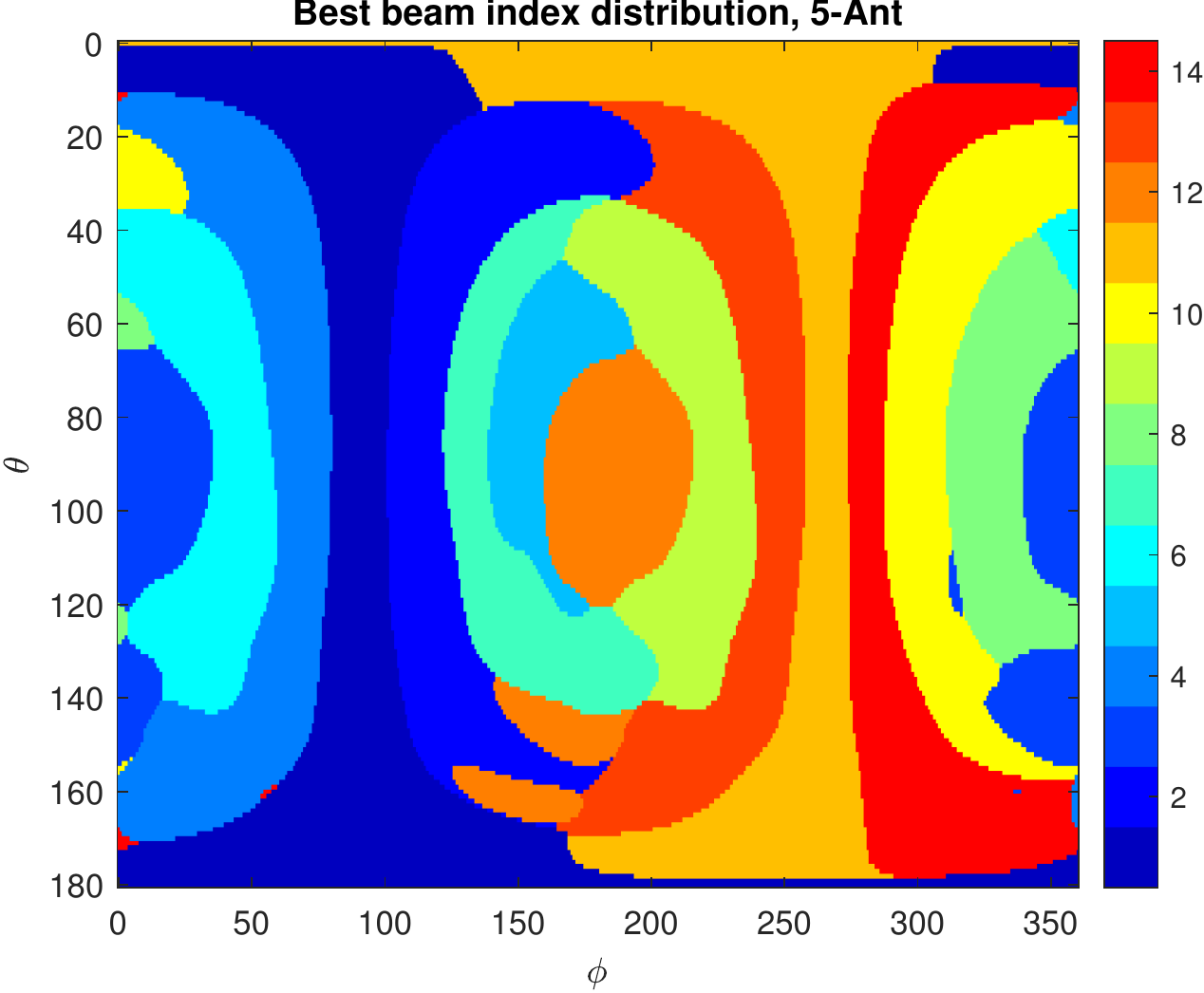}
	}
	\subfigure[4-Ant]{
		\includegraphics[width=0.18 \linewidth]{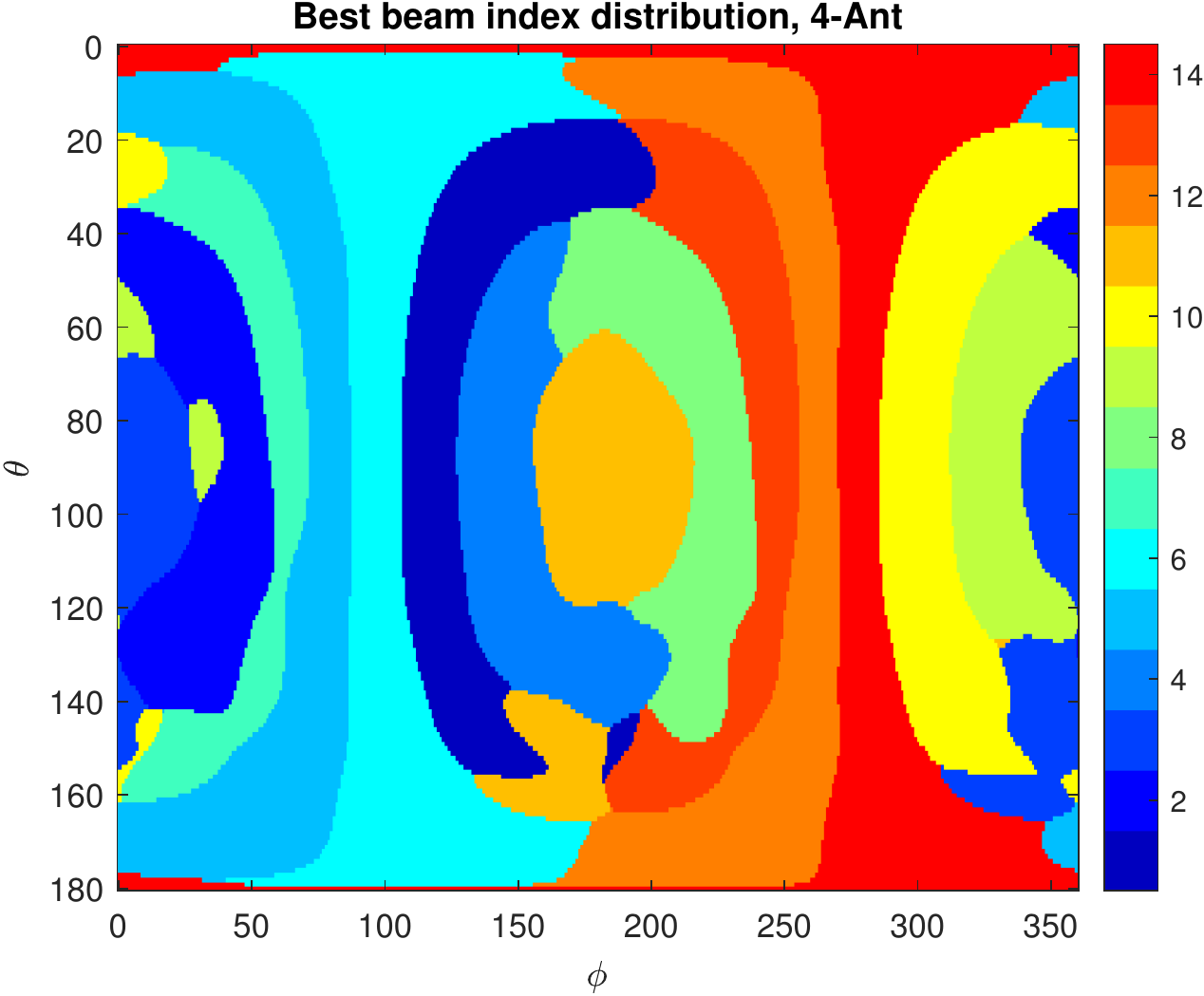}
	}
	\subfigure[3-Ant]{
		\includegraphics[width=0.18 \linewidth]{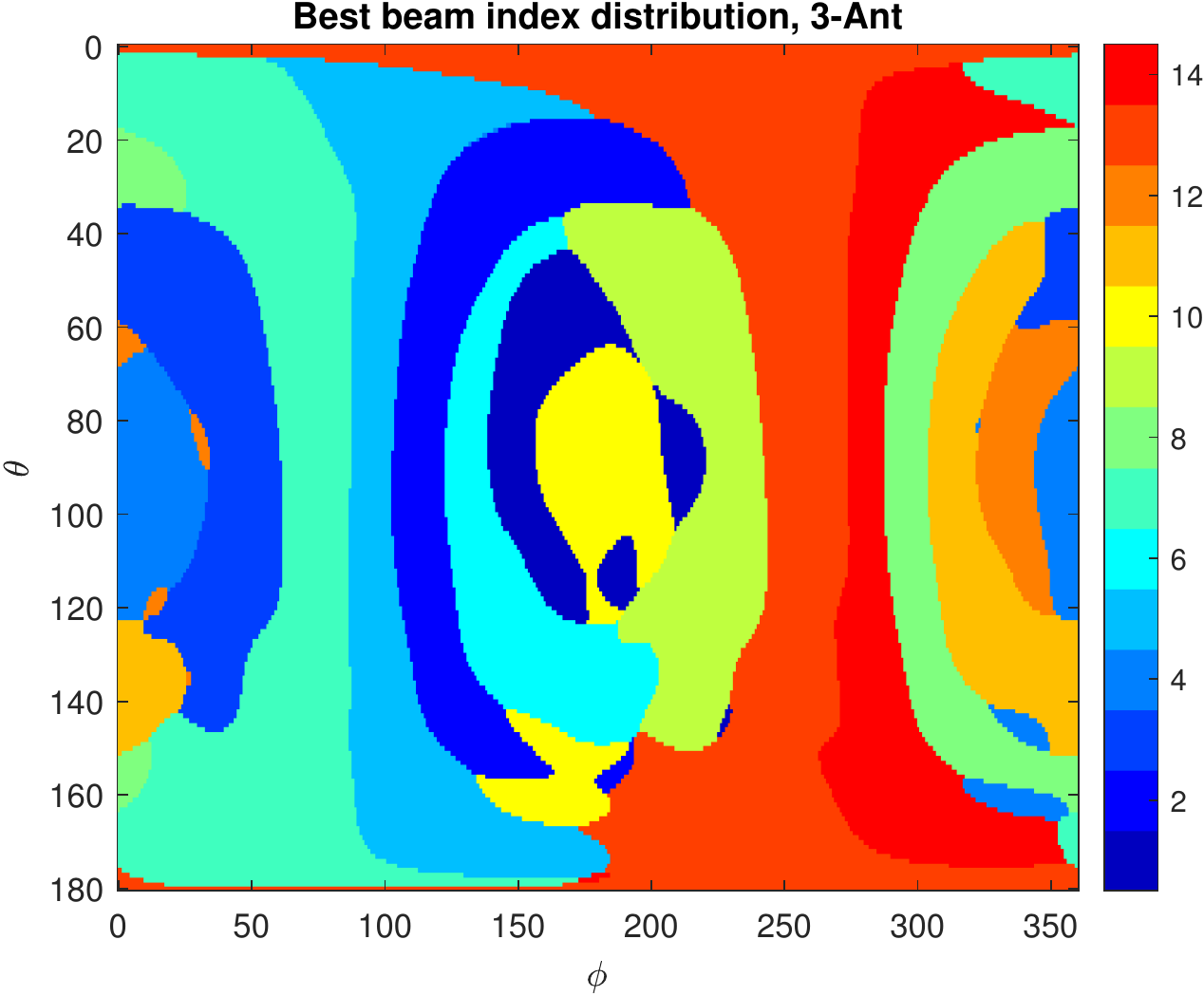}
	}
	\subfigure[2-Ant]{
		\includegraphics[width=0.18 \linewidth]{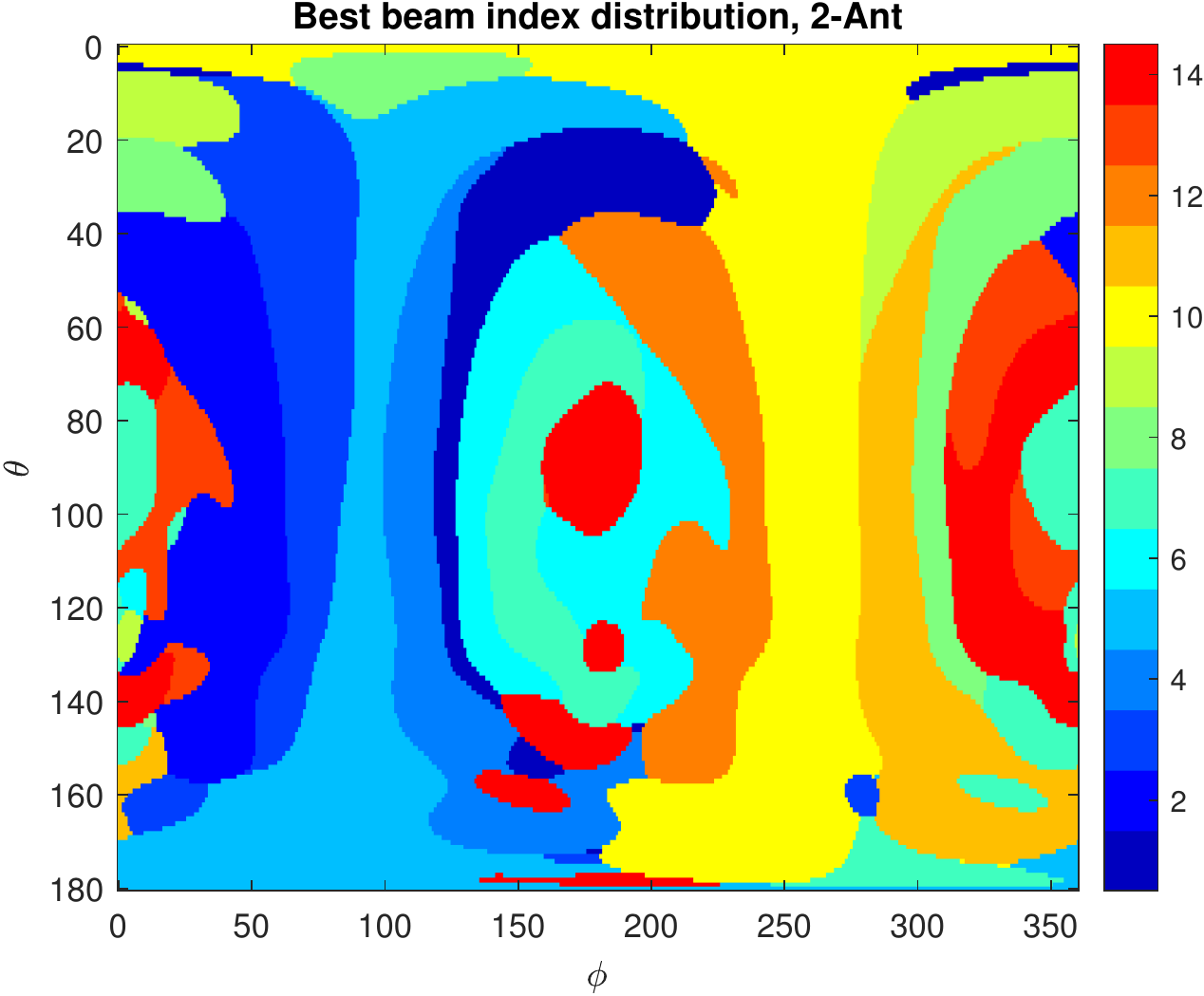}
	}
	\subfigure[1-Ant]{
		\includegraphics[width=0.18 \linewidth]{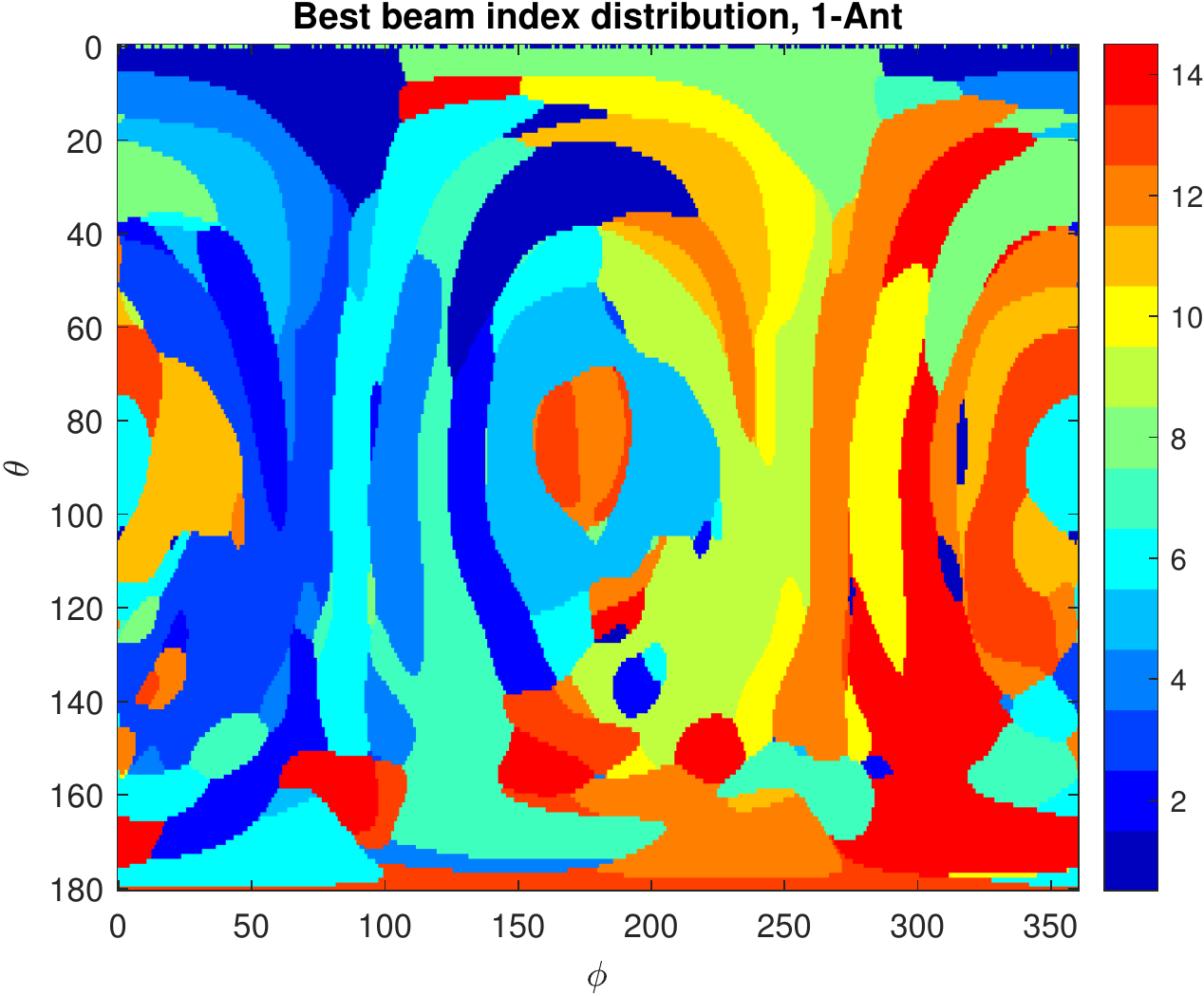}
	}
	\caption{The pattern and the best beam index of the full-chain, and sub-chain codebooks generated by the SC-Max method.}
	\label{fig:SC_Max}
\end{figure*}

\begin{figure*}[th]
	\centering
	\subfigure[5-Ant]{
		\includegraphics[width=0.18 \linewidth]{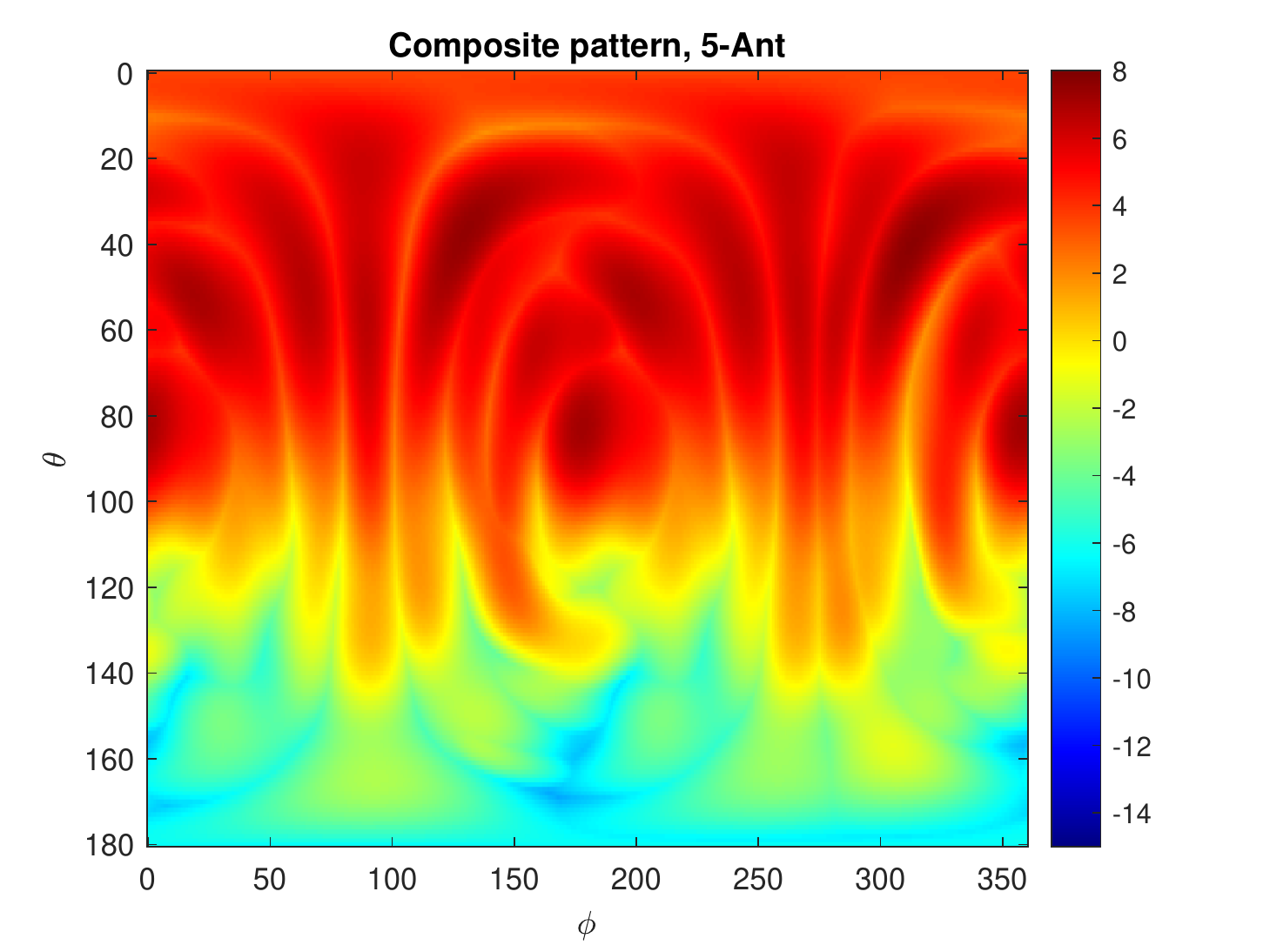}
	}
	\subfigure[4-Ant]{
		\includegraphics[width=0.18 \linewidth]{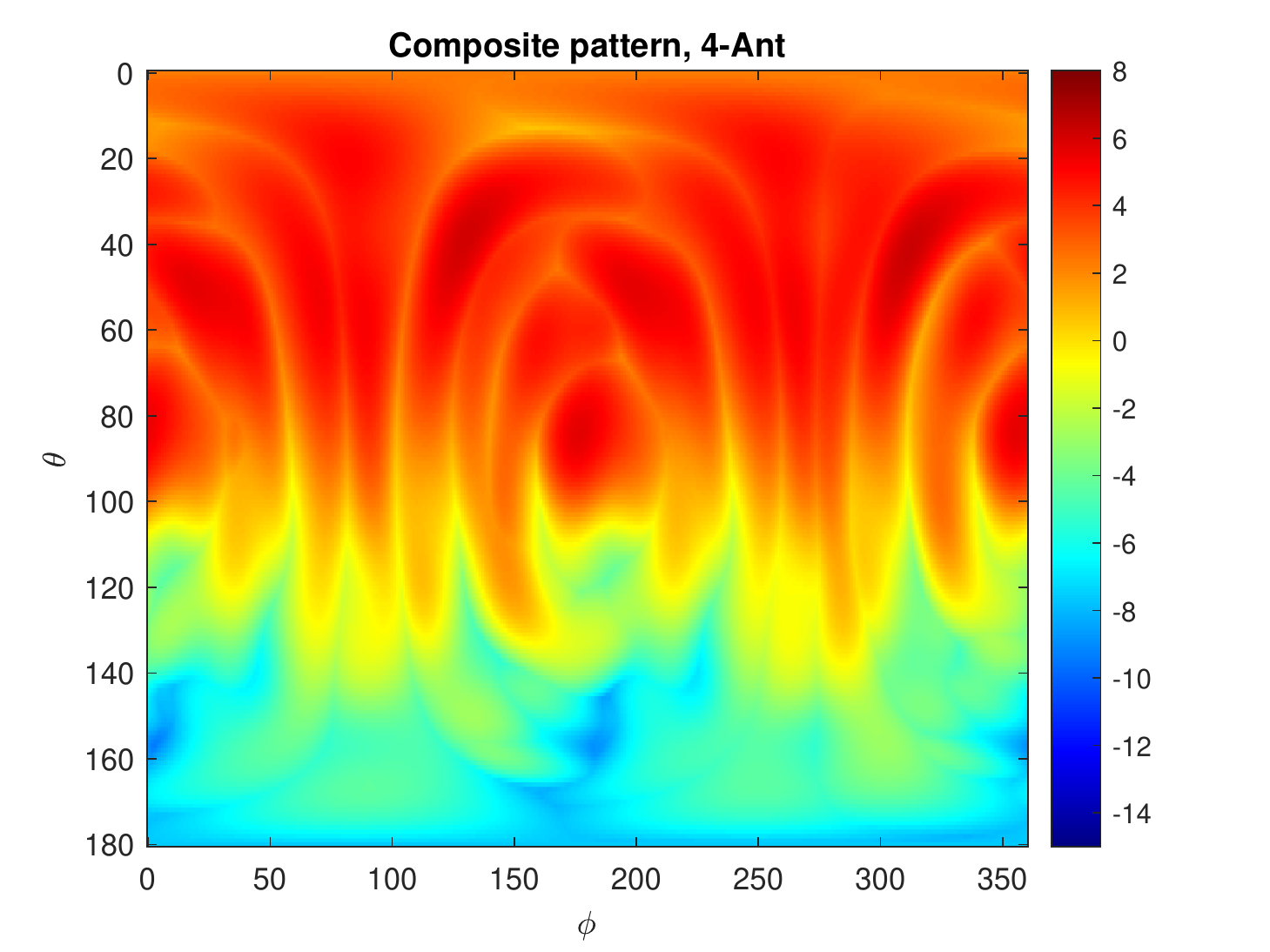}
	}
	\subfigure[3-Ant]{
		\includegraphics[width=0.18 \linewidth]{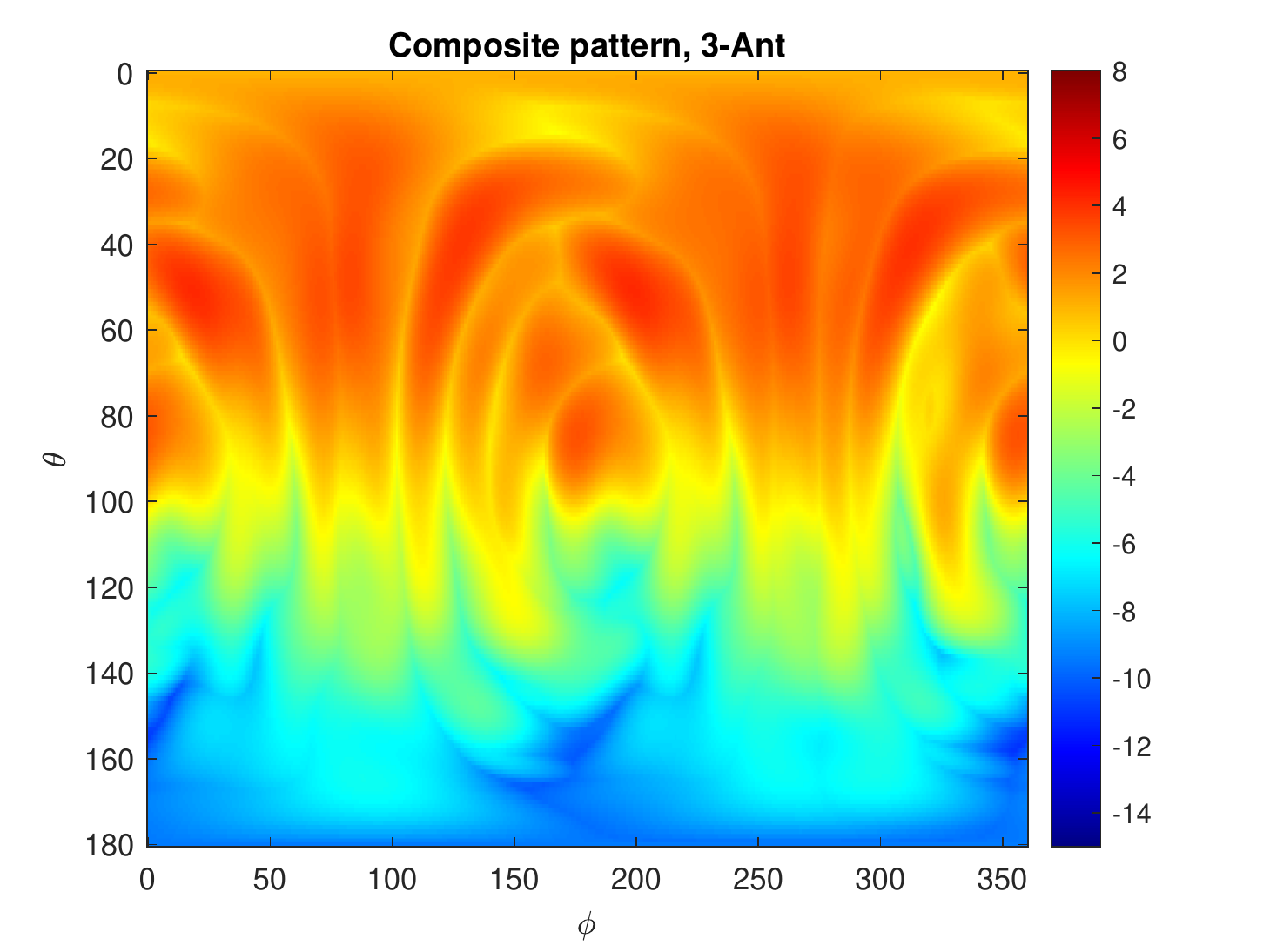}
	}
	\subfigure[2-Ant]{
		\includegraphics[width=0.18 \linewidth]{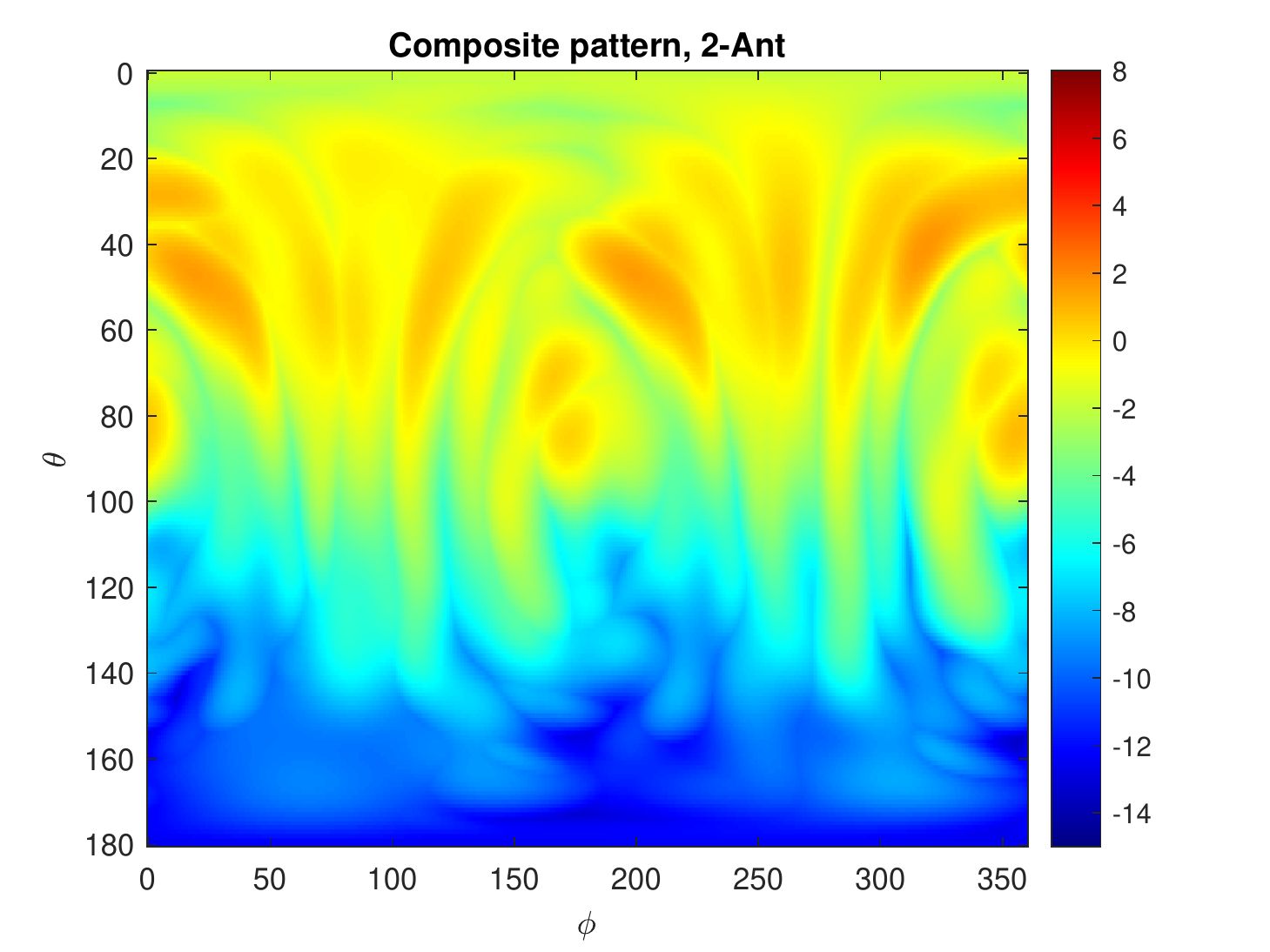}
	}
	\subfigure[1-Ant]{
		\includegraphics[width=0.18 \linewidth]{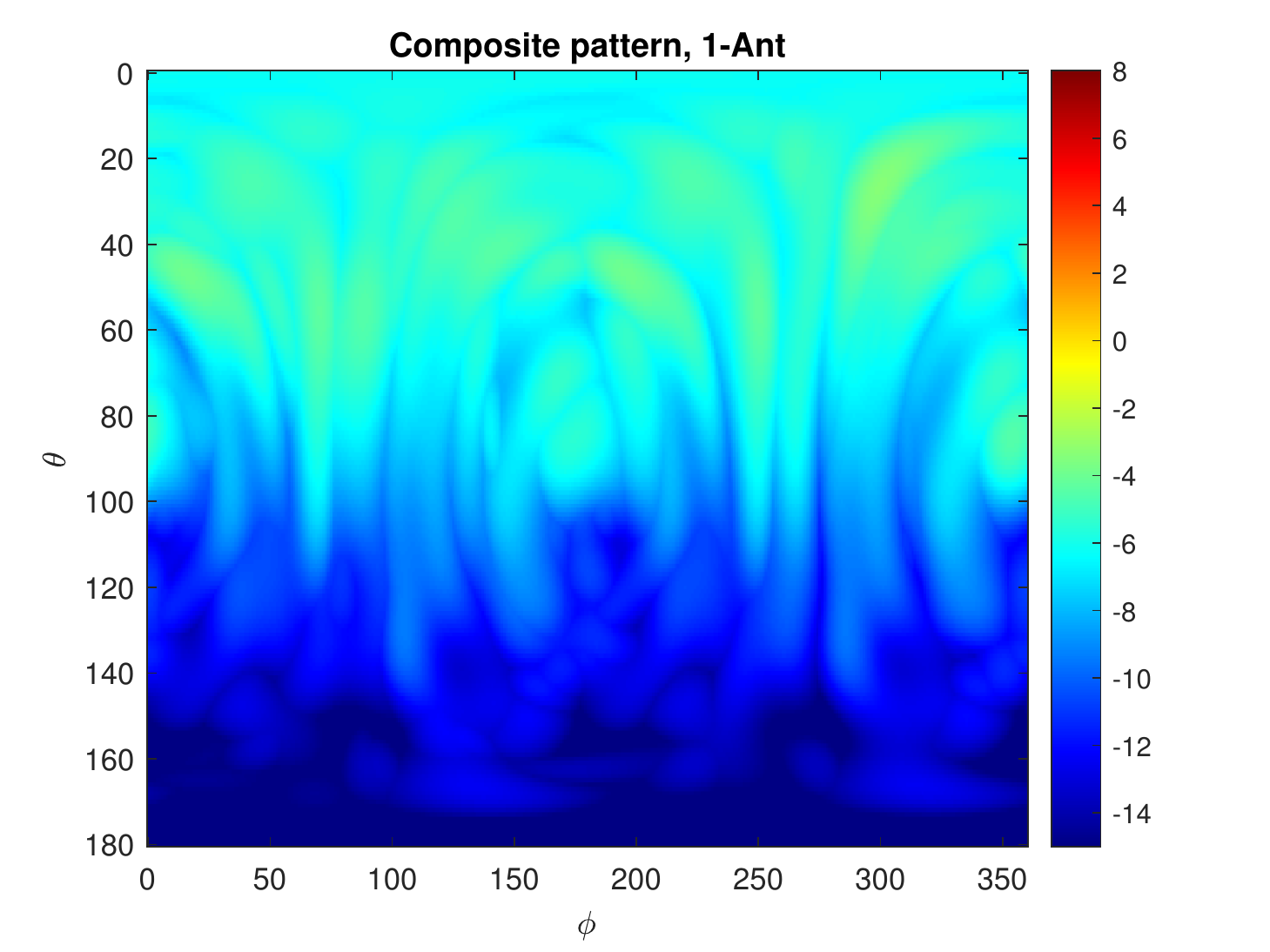}
	}
	\subfigure[5-Ant]{
		\includegraphics[width=0.18 \linewidth]{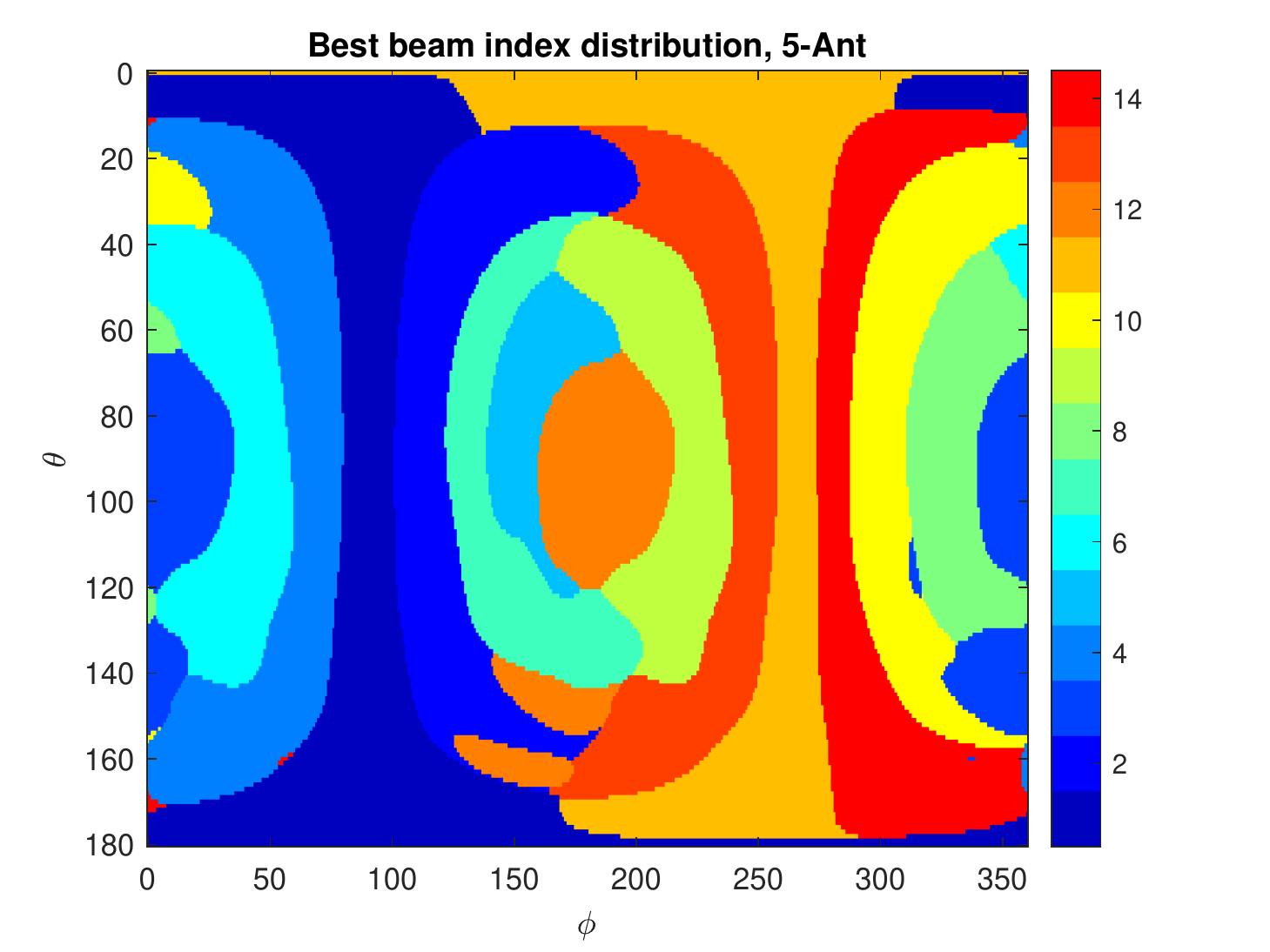}
	}
	\subfigure[4-Ant]{
		\includegraphics[width=0.18 \linewidth]{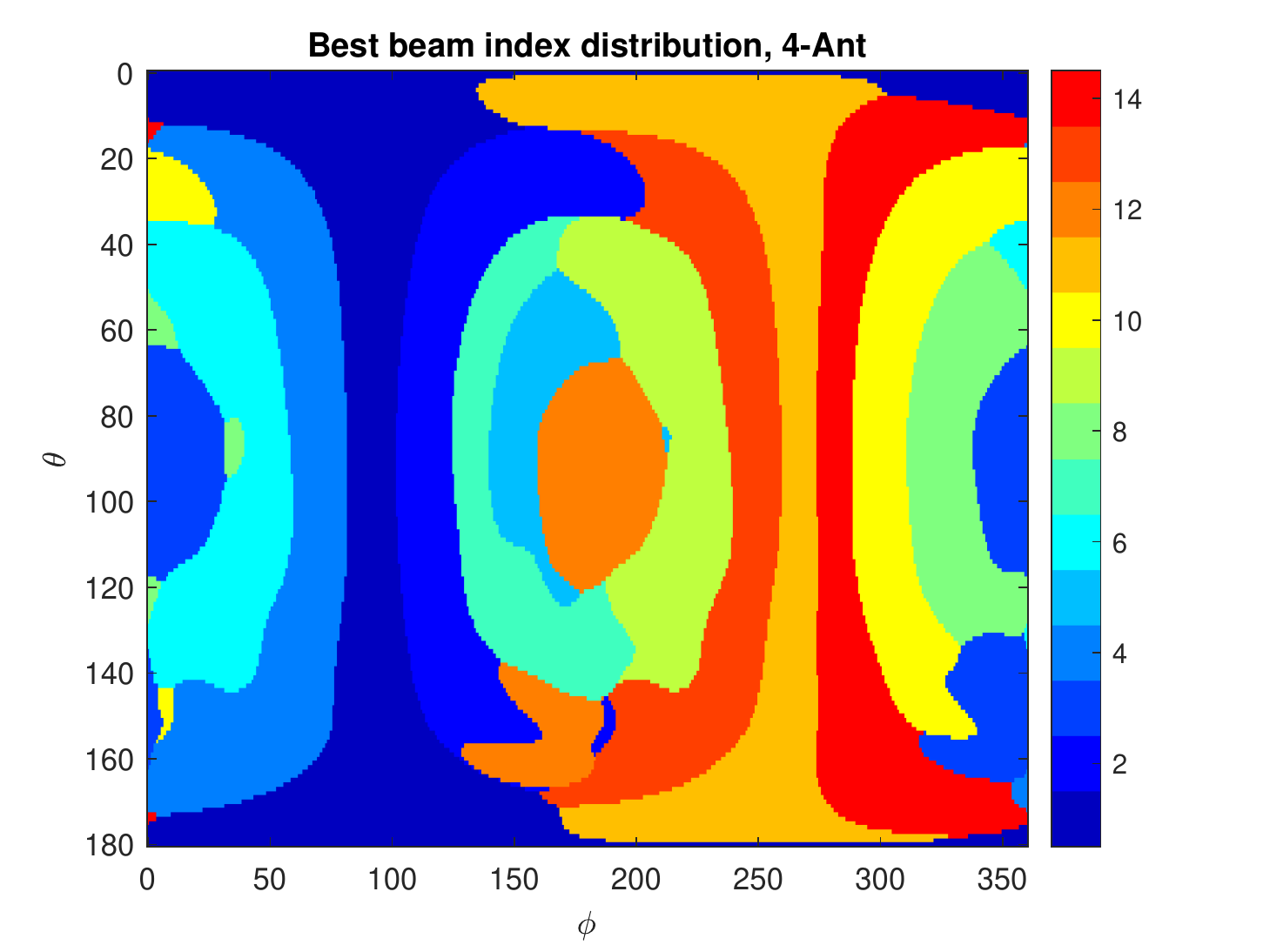}
	}
	\subfigure[3-Ant]{
		\includegraphics[width=0.18 \linewidth]{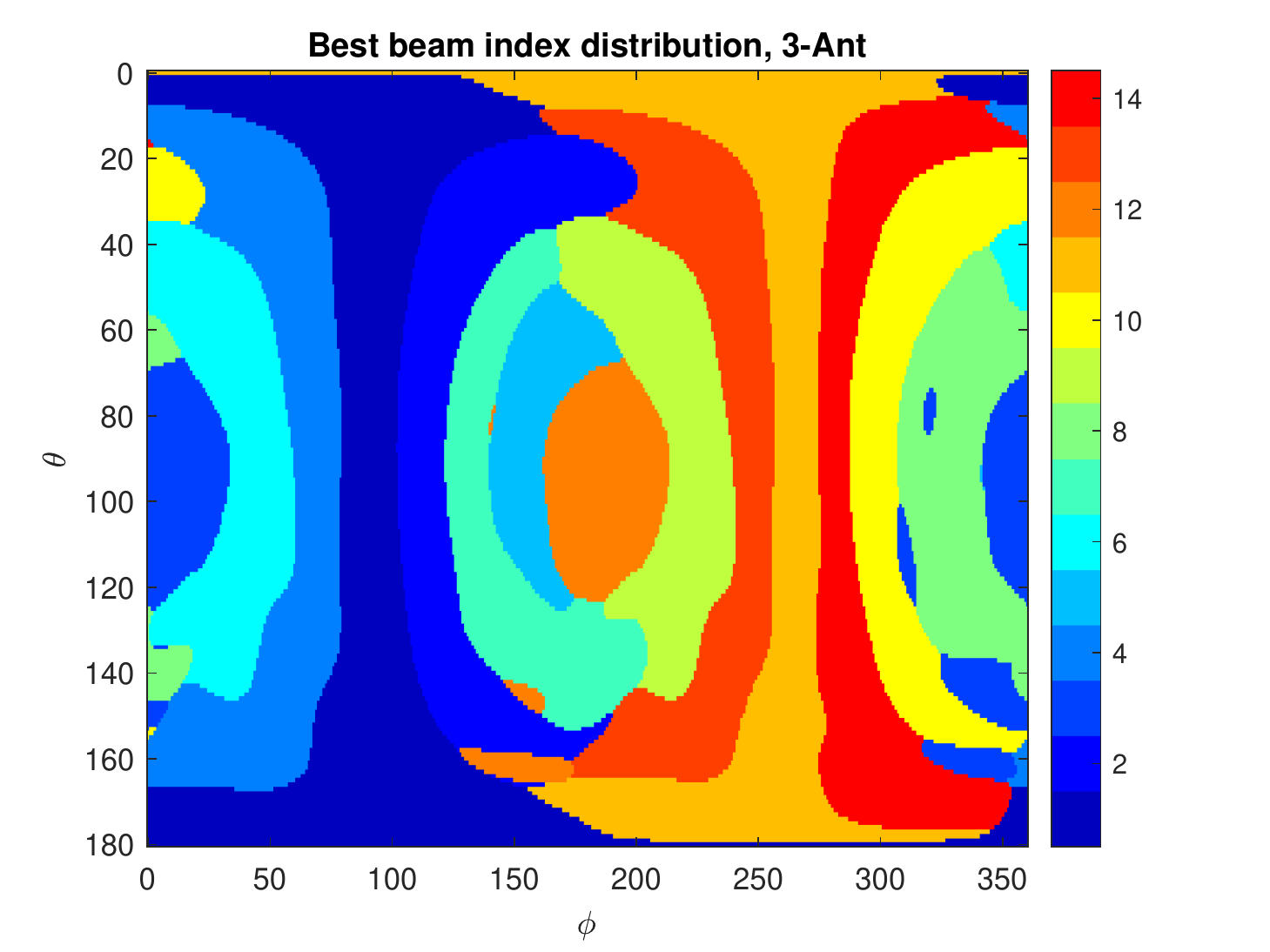}
	}
	\subfigure[2-Ant]{
		\includegraphics[width=0.18 \linewidth]{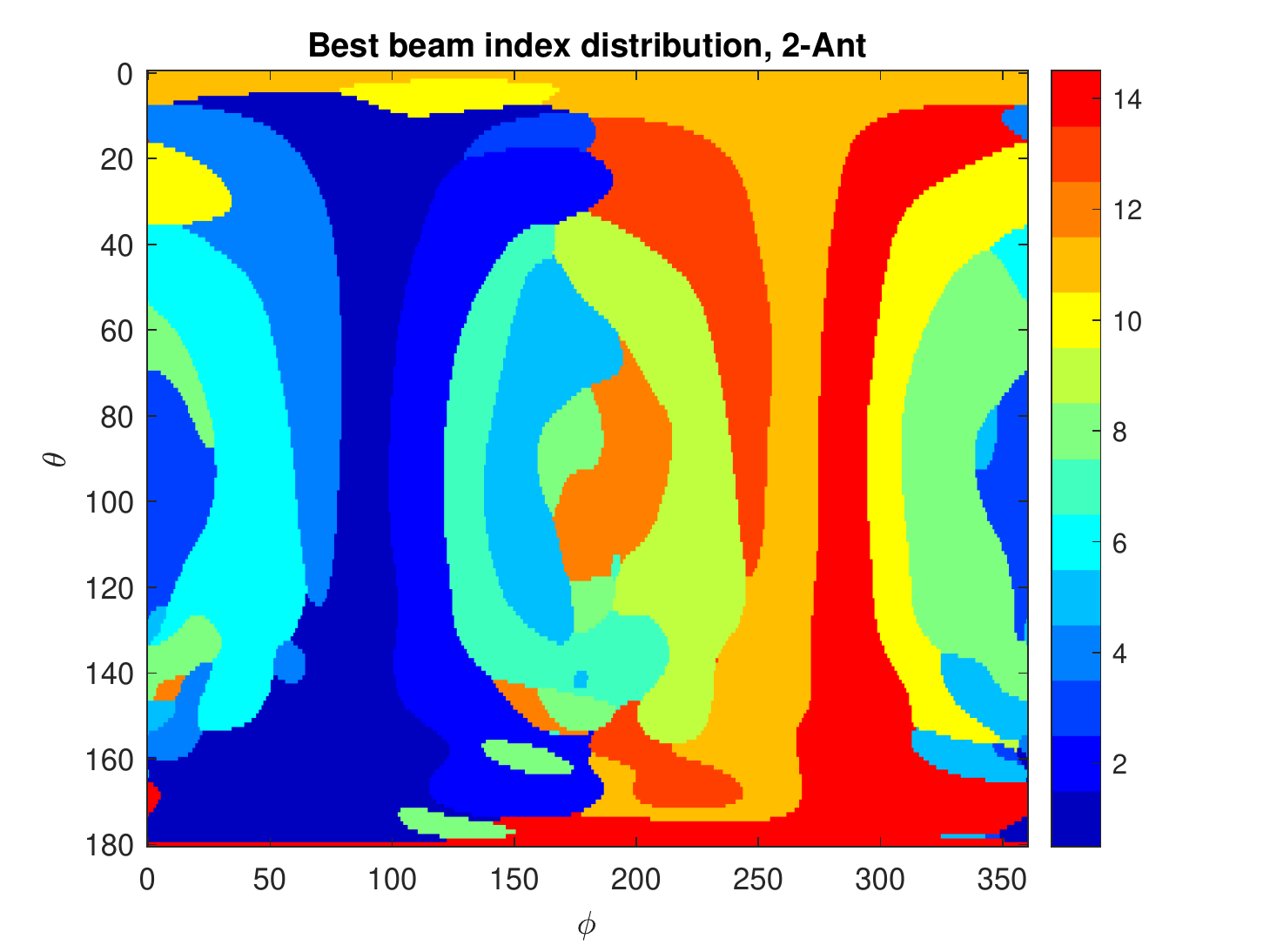}
	}
	\subfigure[1-Ant]{
		\includegraphics[width=0.18 \linewidth]{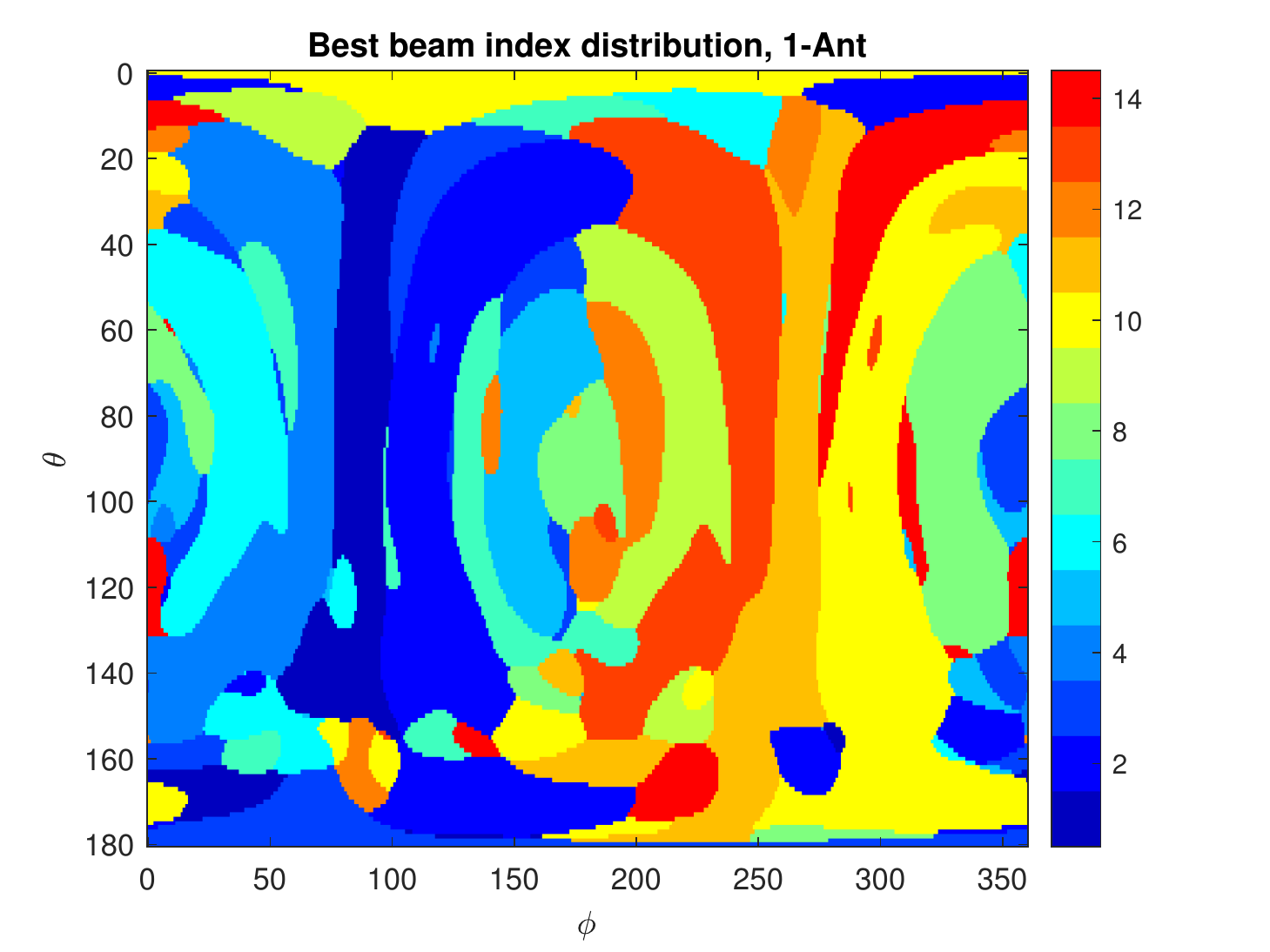}
	}
	\caption{The pattern and the best beam index of the full-chain, and sub-chain codebooks generated by the BC-SC-Max method.}
	\label{fig:BC_SC_Max}
\end{figure*}

\begin{figure*}[th]
	\centering
	\subfigure[Sim-Max]{
		\includegraphics[width=0.31 \linewidth]{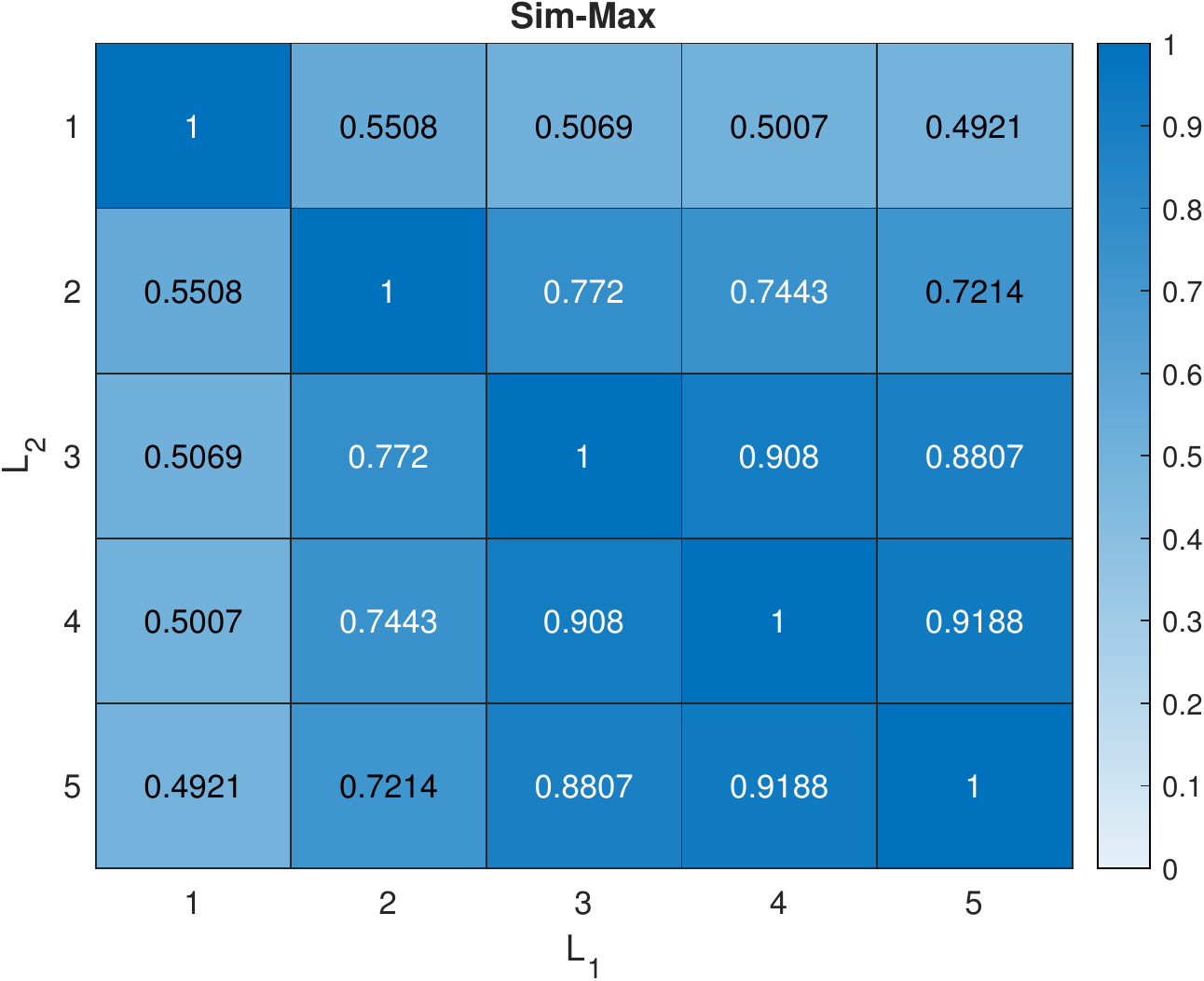}
		\label{fig:Sim_Max_OverlapRatio}}
	\subfigure[SC-Max]{
		\includegraphics[width=0.31 \linewidth]{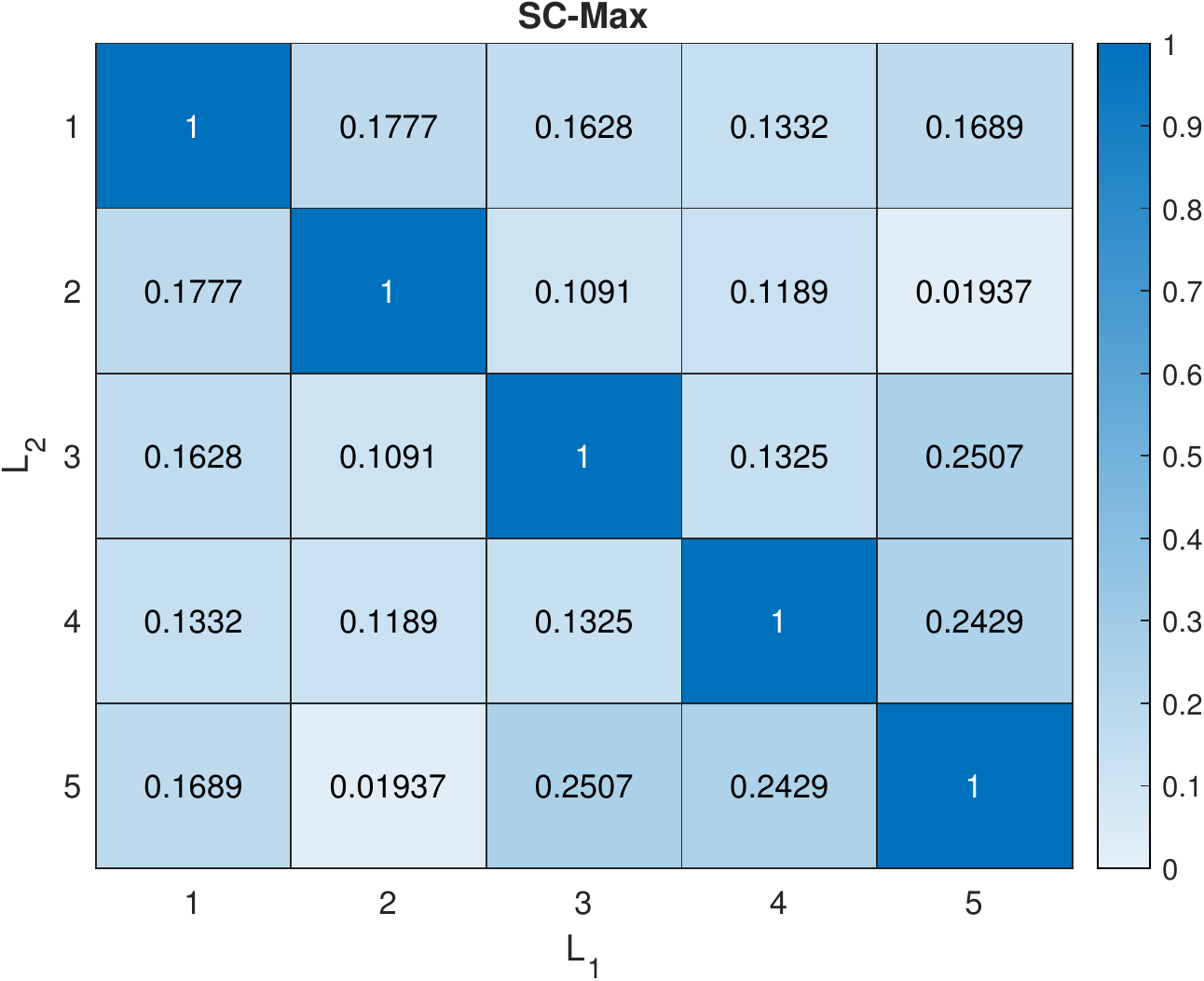}
		\label{fig:SC_Max_OverlapRatio}}
	\subfigure[BC-SC-Max]{
		\includegraphics[width=0.31 \linewidth]{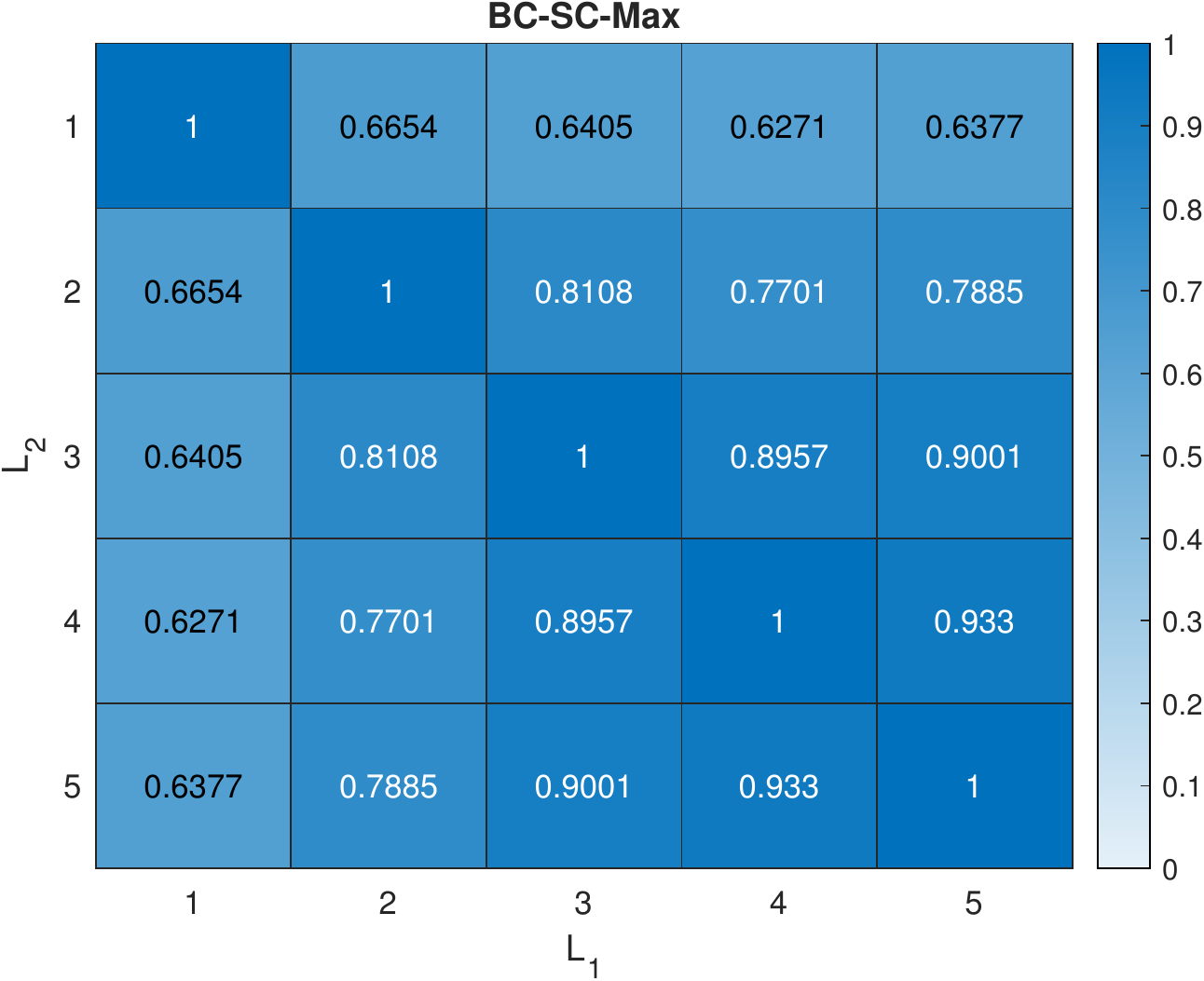}
		\label{fig:BC_SC_Max_OverlapRatio}}
	\caption{The best beam matching rate for different codebooks.}
	\label{fig:OverlapRatio}
\end{figure*}

\begin{figure} [th]
	\centering
	\includegraphics[width=.8\linewidth]{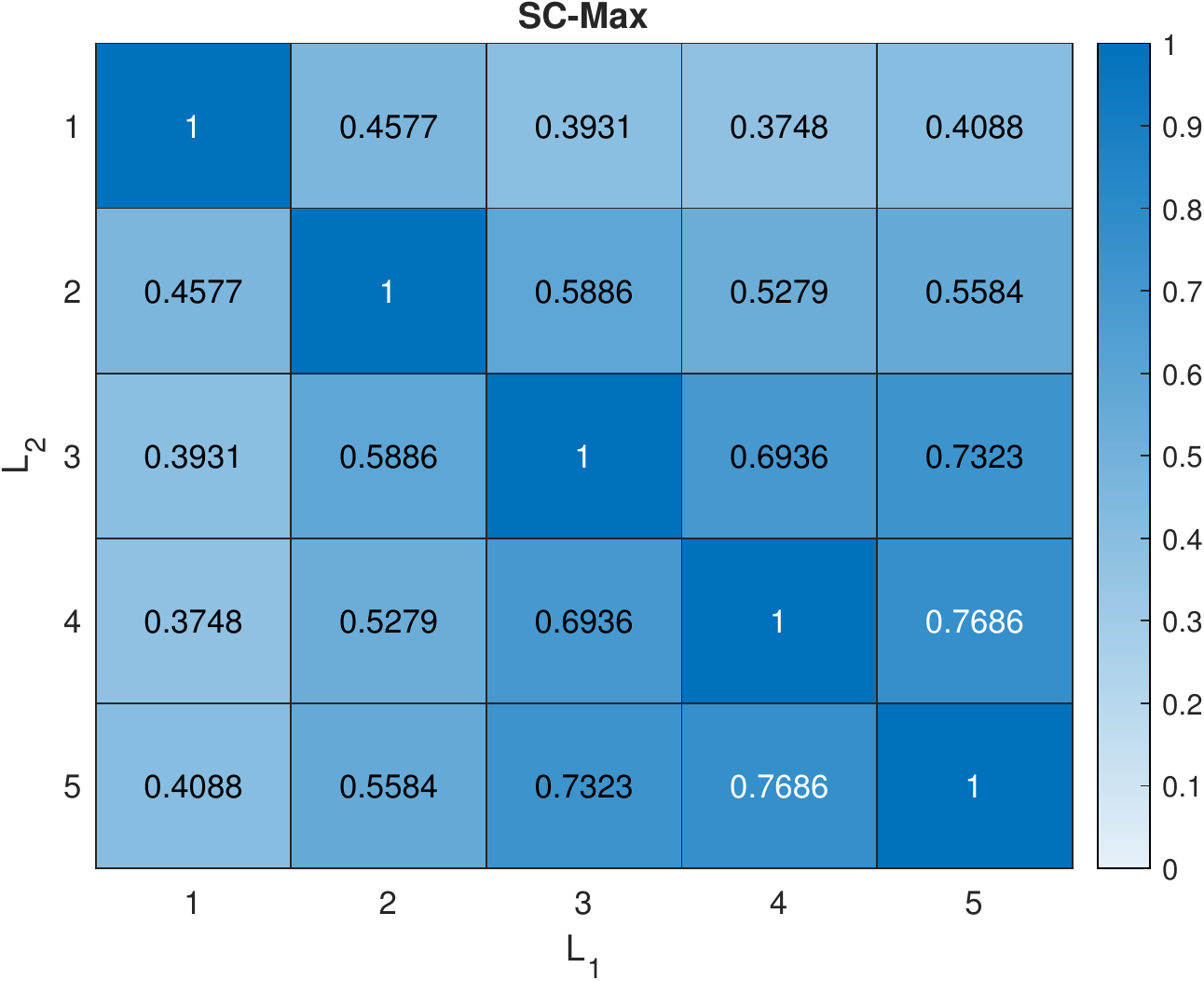}
	\caption{The best beam matching rate of the SC-Max method if the Hungarian algorithm is used to pair the beams.}
	\label{fig:OverlapRatioMatch}
\end{figure}

We simulated a 5G phone with two mmWave 1x5 arrays on the left and right edge by the electromagnetic simulation software. E-field data was generated for each antenna element. Each array has seven beams and the total number of beams is 14. 
The phase shifter resolution is 5-bit. We choose $N_p=10001$ when generating the uniformly distributed sampling points.
The full-chain codebook is generated by the K-Means algorithm \cite{Mo_Jianhua_Access19}.
In the SC-Max method, we initialize the K-Means algorithm with a codebook obtained by a greedy algorithm \cite[Section V]{Mo_Jianhua_Access19}.

\figref{fig:Sim_Max}-\figref{fig:BC_SC_Max} illustrate the generated sub-chain beam codebooks from the Sim-Max, SC-Max and BC-SC-Max method, respectively\footnote{The two linear arrays are both along the y-axis, and their boresight directions are $\l(\theta=90^\circ, \phi=90^\circ\r)$ and $\l(\theta=90^\circ, \phi=270^\circ\r)$, respectively. The region $\theta > 120^\circ$ has lower gain than the other parts because it is slightly blocked by some other components of the phone.}. Subfigure (a)-(e) show the composite beam pattern $\widehat{B}(\theta, \phi) = \max_{1\leq i\leq 14} B_i \left( \theta, \phi \right)$ and Subfigure (f)-(j) illustrate the best beam index distribution $I(\theta, \phi) = \argmax_{1\leq i\leq 14} B_i \left( \theta, \phi \right)$.
The radiation pattern of sub-chain codebooks are weaker than the 5-Ant full-chain codebook, since less antennas are activated.
For the Sim-Max shown in \figref{fig:Sim_Max}, the pattern shapes and the best beam index distributions of the full-chain and sub-chain (especially 4-Ant and 3-Ant) codebooks are more or less similar, which implies the beam correspondence between the full-chain and sub-chain is preserved. The same observation holds for the BC-SC-Max codebooks shown in \figref{fig:BC_SC_Max}. 
In contrast, for the SC-Max method shown in \figref{fig:SC_Max}, there is less similarity between the radiation patterns across the 5 codebooks, and the beam index distributions are also quite different across the 5 codebooks.

We quantify the beam correspondence by checking the probability that the best beam index distribution is same between two codebooks over the unit-sphere.
If the best beam index is same at a location on the unit-sphere, it means that the beam correspondence is preserved for that particular direction, and there is no need to perform another round of beam sweeping when deactivating or activating more antennas. Because of the random UE orientation, we can assume the channel path comes from all the directions with equal probability. Therefore, the proposed metric tells us how often skipping the beam sweeping does not incur performance loss in a single-path channel. We call the metric `best beam matching rate' and define it as,
\begin{align}
	p(L_1, L_2) = \frac{1}{N_p} \sum_{n=1}^{N_p} \mathbbm{1}\Big\{I_{L_1}(\theta_n, \phi_n)=I_{L_2}(\theta_n, \phi_n) \Big\},
\end{align}
where $L_1$, $L_2$ are the number of activated antennas of two codebooks.
The proposed metric for the three methods is shown in \figref{fig:OverlapRatio}. First, we find that for the Sim-Max and BC-SC-Max codebooks, it is quite safe to switch among 5-Ant, 4-Ant and 3-Ant codebooks. The beam correspondence is preserved for more than $90\%$ of the time. Second, BC-SC-Max is much better than Sim-Max when switching between 1-Ant codebook and another codebook. 
Third, SC-Max codebooks have very low matching rate ($\leq 25\%$) because there is no consideration of beam mapping at all in the design procedure. 

To improve the matching rate of SC-Max codebooks, we repair the beams rather than simply pairing the beams with same index. The procedure is as follows. We first identify the dominate sampling points of each beam,
\begin{align}
	\mathcal{D}_{k, L_A} = \left\{(\theta_n, \phi_n) \bigg| k=\argmax_{1\leq i\leq K} B_{i, L_A}(\theta_n, \phi_n) \right\}. 
\end{align}
We then find the intersection between two beams from different codebooks,
\begin{align}
\mathcal{C}_{L_1, L_2}(i, j) = \mathcal{D}_{i, L_1} \cap \mathcal{D}_{j, L_2}. 
\end{align}
The cardinality of the intersection $|\mathcal{C}_{L_1, L_2}(i, j)|$ is treated as the benefit of pairing Beam $i$ from $L_1$-Ant codebook and Beam $j$ from $L_2$-Ant codebook. Then Hungarian algorithm is applied to find the best pairing maximizing the total benefit. \figref{fig:OverlapRatioMatch} shows the best beam matching rate after repairing. It is much better than the previous results shown in \figref{fig:SC_Max_OverlapRatio}, but is still clearly worse than the Sim-Max and BC-SC-Max method in \figref{fig:Sim_Max_OverlapRatio} and \figref{fig:BC_SC_Max_OverlapRatio}. It implies that repairing is not a sufficient solution to maintain the beam correspondence.


\begin{figure} [th]
	\centering
	\includegraphics[width=1.0\linewidth]{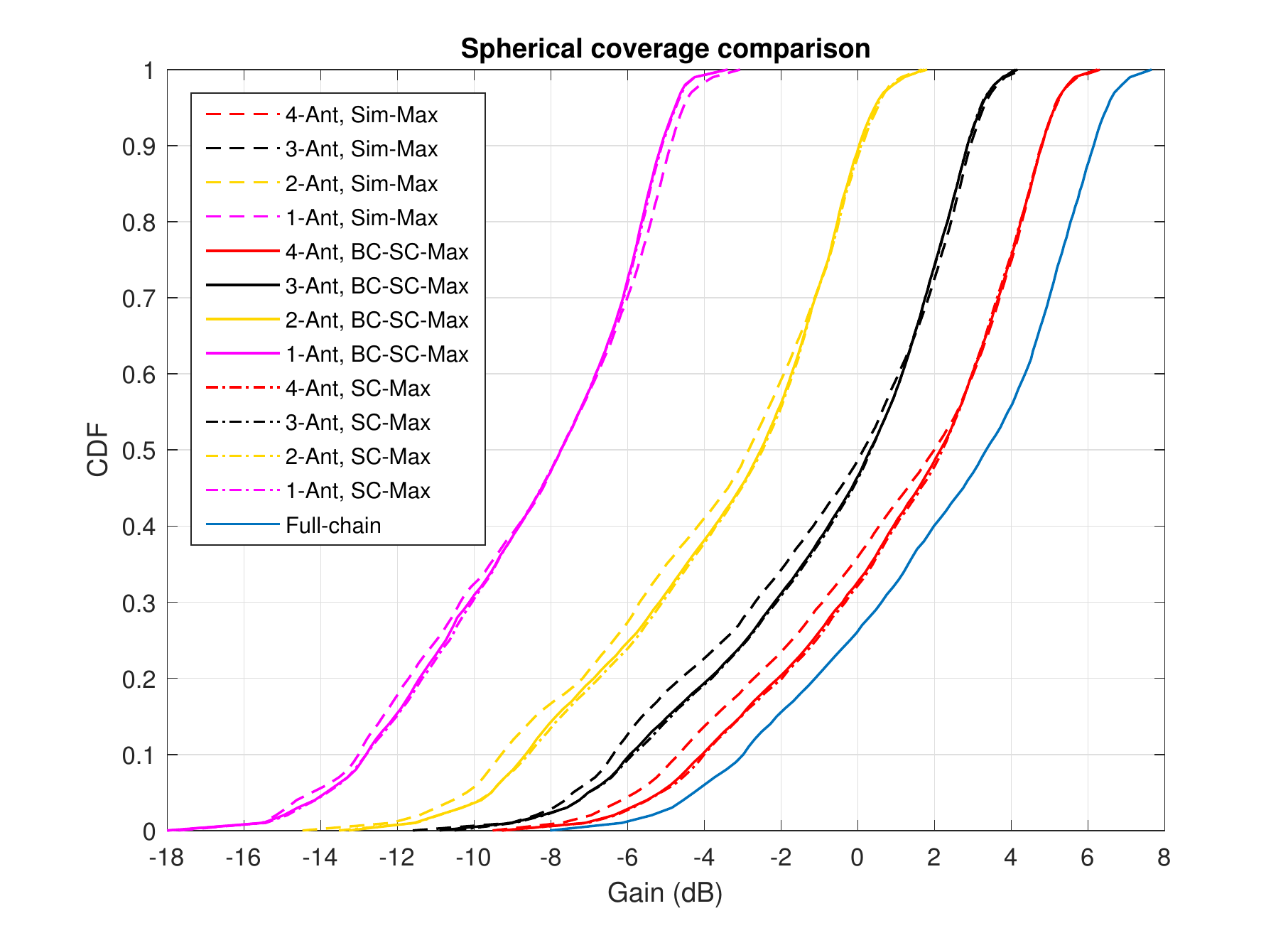}
	\caption{The spherical coverage comparison of the sub-chain beam codebooks}
	\label{fig:SC_CDF_compare}
\end{figure}

Last but not least, we compare the spherical coverage of those codebooks by checking the composite beam gain CDF on the unit-sphere. \figref{fig:SC_CDF_compare} shows the CDF curves for the full-chain, and sub-chain codebooks of the three methods. Note that the repairing does not change the spherical coverage of the SC-Max codebooks. First, as expected, the spherical coverage improves with the number of antennas and the 5-Ant case is the best one. Second, the spherical coverage of the Sim-Max codebooks are worse than the other two methods, especially in the low percentile region. Third, we find that the BC-SC-Max method achieves the similar spherical coverage as the SC-Max method!
 
\section{Conclusion}

In this paper, we proposed a practical beam operation scheme for mmWave 5G devices to increase the utilization of mmWave band. Because of the high power consumption and heating of mmWave antennas, mmWave devices frequently fall back to sub-6 GHz band, and sporadically utilize the mmWave band. We proposed a sub-chain beam operation which deactivates part of the mmWave antenna array in the uplink transmission when the device is overheating. The antenna deactivation, however, could destroy the downlink-uplink beam correspondence. The more the antenna deactivation, the worse the beam correspondence.
We proposed three methods to carefully design the sub-chain codebooks. The Sim-Max method generates sub-chain beams resembling the shape of full-chain beams, the SC-Max method optimizes the spherical coverage of the sub-chain codebooks, and the BC-SC-Max method takes into account both the similarity and spherical coverage.

We performed extensive simulations of a real 5G phone to compare the performance of the three methods. We found that the BC-SC-Max method is the best one. It can achieve superior spherical coverage close to the SC-Max method which is dedicated to spherical coverage maximization. Meanwhile, it maximally maintains the beam correspondence. The beam correspondence holds for $90\%$ of the time when switching among full-chain, 4-Ant and 3-Ant sub-chain beam codebooks. When UE chooses to use 1-Ant beam to save the power as much as possible, the BC-SC-Max method still preserves the beam correspondence for more than $60\%$ of the time.

\bibliographystyle{IEEEtran}
\bibliography{IEEEabrv,BeamBook}
\end{document}

%% file: input.tex
\usepackage{acronym}
\usepackage{amsfonts}
\usepackage{bbm}
\usepackage{times}
\usepackage{cite}
\usepackage{amsmath}
\usepackage{array}
\usepackage{amssymb}
\usepackage{stfloats}
\usepackage{diagbox}
\usepackage{graphicx}
\usepackage{footnote}
\usepackage{amsthm}
\usepackage{booktabs}
\usepackage{array}
\usepackage{algorithmic}
\usepackage{algorithm}
\usepackage{subeqnarray}
\usepackage{cases}
\usepackage{threeparttable}
\usepackage{color}
\usepackage{epstopdf}
\usepackage{multirow}
\usepackage{tabularx}
\usepackage{enumerate}
\usepackage{multicol}
\usepackage{subfigure}

\newcommand{\figref}[1]{{Fig.}~\ref{#1}}

\def\bb0{{\mathbb{0}}}

\def\ba{{\mathbf{a}}}
\def\bb{{\mathbf{b}}}

\def\bw{{\mathbf{w}}}

\def\b0{{\mathbf{0}}}

\def\bM{{\mathbf{M}}}

\def\sf0{{\mathsf{0}}}

\def\rm0{{\mathrm{0}}}

\def\l{\left}
\def\r{\right}

\acrodef{CSI}[CSI]{channel state information}
\acrodef{CSIT}[CSIT]{channel state information at the transmitter}
\acrodef{CSIR}[CSIR]{channel state information at the receiver}
\acrodef{MIMO}[MIMO]{multiple-input multiple-output}
\acrodef{SISO}[SISO]{single-input single-output}
\acrodef{MISO}[MISO]{multiple-input single-output}
\acrodef{SIMO}[SIMO]{single-input multiple-output}
\acrodef{ADCs}[ADCs]{analog-to-digital convertors}
\acrodef{SNR}[SNR]{signal-to-noise ratio}
\acrodef{AWGN}[AWGN]{additive white Gaussian noise}
\acrodef{MRT}[MRT]{maximal ratio transmission}
\acrodef{DFT}[DFT]{Discrete Fourier Transform}
\acrodef{ULA}[ULA]{uniform linear array}
\acrodef{UPA}[UPA]{uniform planar array}
\acrodef{LS}[LS]{least squares}
\acrodef{ML}[ML]{maximum likelihood}
\acrodef{AML}[AML]{approximate ML}
\acrodef{LMMSE}[LMMSE]{linear MMSE}
\acrodef{ALMMSE}[ALMMSE]{approximate LMMSE}
\acrodef{QIHT}[QIHT]{quantized iterative hard thresholding}
\acrodef{QIST}[QIST]{quantized iterative soft thresholding}
\acrodef{PAPR}[PAPR]{peak-to-average power ratio}
\acrodef{SVD}[SVD]{singular value decomposition}
\acrodef{CDF}[CDF]{cumulative distribution function}
\acrodef{EIRP}[EIRP]{equivalent isotropic radiated power}
\acrodef{HFSS}[HFSS]{high-frequency structure simulator}
\acrodef{QCQP}[QCQP]{quadratically constrained quadratic program}
\acrodef{SDR}[SDR]{semi-definite relaxation}
\acrodef{SDP}[SDP]{semi-definite programming}
\acrodef{KKT}[KKT]{Karush-Kuhn-Tucker}

\DeclareMathOperator*{\argmax}{argmax} 

